\documentclass[aps,pra,twocolumn,superscriptaddress]{revtex4-1}  
\usepackage{graphicx,color}  
\usepackage{dcolumn}   
\usepackage{bm}        
\usepackage{amssymb}   
\usepackage{subfigure}
\usepackage{lipsum}   

\usepackage[heightadjust=all,valign=t]{floatrow}
\usepackage{fr-subfig}
\setcitestyle{super}

\usepackage{sidecap}
\usepackage{comment}
\usepackage{mathrsfs}
\usepackage{soul}
\usepackage{comment} 
\usepackage{hyperref}
\usepackage{xcolor}

\newcommand{\beginsupplement}{%
    \setcounter{table}{0}
    \renewcommand{\thetable}{S\arabic{table}}%
    \setcounter{figure}{0}
    \renewcommand{\thefigure}{S\arabic{figure}}%
}

\providecommand{\ignore}[1]{}

\newcommand{\ket}[1]{|#1\rangle}

\newcommand{\expect}[1]{\langle #1\rangle}
\def\lsim{\mathrel{\rlap{\lower4pt\hbox{\hskip1pt$\sim$}}
    \raise1pt\hbox{$<$}}}                
\def\gsim{\mathrel{\rlap{\lower4pt\hbox{\hskip1pt$\sim$}}
    \raise1pt\hbox{$>$}}}                

\begin{document}

\title{Correlated Charge Noise and Relaxation Errors in Superconducting Qubits}

\author{C. D. Wilen} 
\affiliation{Department of Physics, University of Wisconsin-Madison, Madison, Wisconsin 53706, USA}

\author{S. Abdullah} 
\affiliation{Department of Physics, University of Wisconsin-Madison, Madison, Wisconsin 53706, USA}

\author{N. A.~Kurinsky} 
\affiliation{Fermi National Accelerator Laboratory, Center for Particle Astrophysics, Batavia, IL 60510 USA} 
\affiliation{Kavli Institute for Cosmological Physics, University of Chicago, Chicago, IL 60637, USA}

\author{C. Stanford}
\affiliation{Department of Physics, Stanford University, Stanford, CA 94305 USA}

\author{L. Cardani} 
\affiliation{INFN Sezione di Roma, P.le Aldo Moro 2, 00185, Rome, Italy}

\author{G. D'Imperio} 
\affiliation{INFN Sezione di Roma, P.le Aldo Moro 2, 00185, Rome, Italy}

\author{C. Tomei} 
\affiliation{INFN Sezione di Roma, P.le Aldo Moro 2, 00185, Rome, Italy}

\author{L. Faoro} 
\affiliation{Department of Physics, University of Wisconsin-Madison, Madison, Wisconsin 53706, USA}
\affiliation{Sorbonne Universite, Laboratoire de Physique Theorique et Hautes Energies, UMR 7589 CNRS, Tour 13, 5\`eme Etage, 4 Place Jussieu, F-75252 Paris 05, France}

\author{L. B.~Ioffe} 
\affiliation{Google, Inc., Venice, CA 90291 USA}

\author{C. H. Liu} 
\affiliation{Department of Physics, University of Wisconsin-Madison, Madison, Wisconsin 53706, USA}

\author{A. Opremcak} 
\affiliation{Department of Physics, University of Wisconsin-Madison, Madison, Wisconsin 53706, USA}

\author{B. G. Christensen} 
\affiliation{Department of Physics, University of Wisconsin-Madison, Madison, Wisconsin 53706, USA}

\author{J. L.~DuBois} 
\affiliation{Physics Division, Lawrence Livermore National Laboratory, Livermore, California 94550, USA}

\author{R. McDermott}
\email[]{rfmcdermott@wisc.edu}
\affiliation{Department of Physics, University of Wisconsin-Madison, Madison, Wisconsin 53706, USA}

\date{\today}

\begin{abstract}
The central challenge in building a quantum computer is error correction. Unlike classical bits, which are susceptible to only one type of error, quantum bits (“qubits”) are susceptible to two types of error, corresponding to flips of the qubit state about the $X$- and $Z$-directions. While the Heisenberg Uncertainty Principle precludes simultaneous monitoring of $X$- and $Z$-flips on a single qubit, it is possible to encode quantum information in large arrays of entangled qubits that enable accurate monitoring of all errors in the system, provided the error rate is low. Another crucial requirement is that errors cannot be correlated.  Here, we characterize a superconducting multiqubit circuit and find that charge fluctuations are highly correlated on a length scale over 600~$\mu$m; moreover, discrete charge jumps are accompanied by a strong transient suppression of qubit energy relaxation time across the millimeter-scale chip. The resulting correlated errors are explained in terms of the charging event and phonon-mediated quasiparticle poisoning associated with absorption of gamma rays and cosmic-ray muons in the qubit substrate. Robust quantum error correction will require the development of mitigation strategies to protect multiqubit arrays from correlated errors due to particle impacts.
\end{abstract}

\maketitle

The two-dimensional surface code is widely seen as a promising approach to realization of a fault-tolerant quantum computer based on superconducting integrated circuits \cite{Fowler12}. In this architecture, quantum information is encoded in a two-dimensional fabric of superconducting qubits with nearest-neighbor connectivity. Provided that gate operations and measurements are performed above a certain \textit{fault-tolerant threshold}, it is possible to uniquely identify and correct errors in the system by monitoring multiqubit parity operators of the form $XXXX$ and $ZZZZ$, where $X$ and $Z$ are single-qubit Pauli operators. In recent years, a number of groups have achieved beyond-threshold fidelities for single- and two-qubit gate operations \cite{Barends14,Sheldon2016} and for qubit measurement \cite{Jeffrey14,Walter17,Opremcak20}, and steady improvements in performance are expected. The rigorous proof that it is possible, in principle, to achieve fault tolerance once threshold levels of fidelity are reached underpins much of the optimism for the surface code. However, this proof rests on the assumption that errors across the multiqubit array are uncorrelated in both space and time. While it is possible to mitigate errors that are weakly correlated across neighboring qubits \cite{Faoro20}, quantum error correction will break down in the face of simultaneous errors that are correlated over large length scales.

In this article, we demonstrate spatially correlated charge fluctuations in a superconducting multiqubit chip over length scales of hundreds of microns, accompanied by correlated relaxation errors that extend over several millimeters. The data are compatible with absorption in the qubit substrate of cosmic-ray muons and $\gamma$-rays from background radioactivity. We perform detailed numerical modeling to determine the spatial profile of the charge burst associated with an absorption event; in addition, we present a simple model that describes the propagation of energy released by the event via scattering of pair-breaking phonons. These results have far-reaching implications for proposed error correction schemes such as the surface code that rely on large-scale arrays of qubits to monitor multiqubit stabilizers. A thorough understanding of the physics of particle impact events will be required to develop appropriate mitigation strategies and to engineer new approaches for fault-tolerant multiqubit arrays. 

\begin{figure}[ht]
\includegraphics[width=\columnwidth]{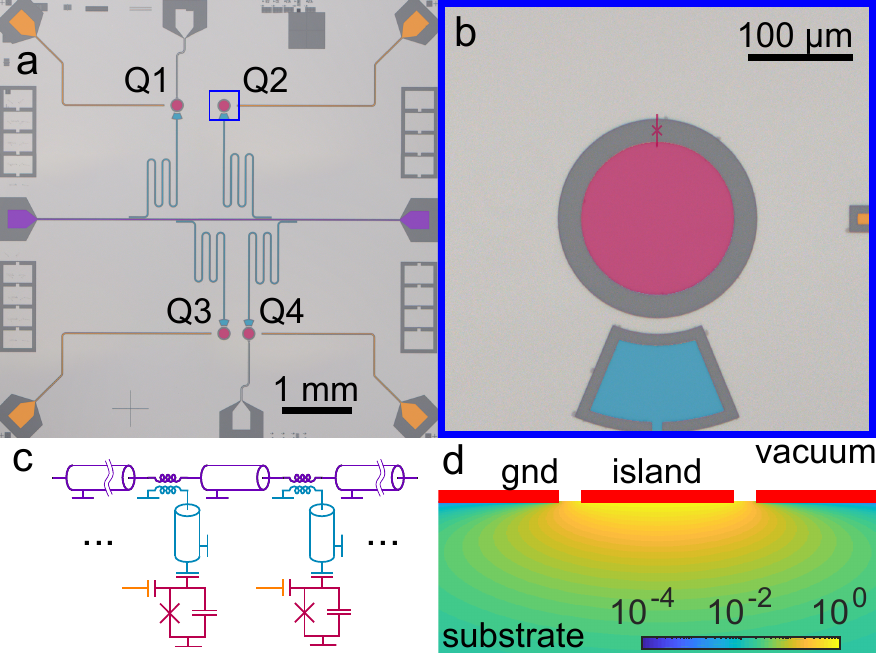}
\caption{ \textbf{Chip layout and charge response.} (\textbf{a}) Optical micrograph of the multiqubit chip. Four charge-sensitive transmon qubits (magenta) are coupled to local readout resonators (cyan) and charge gate lines (orange). The readout resonators are coupled to a common feedline (purple). 
(\textbf{b}) Closeup view of a single qubit.  (\textbf{c}) Circuit diagram of the chip. Color coding matches the false coloring in parts (a) and (b).  (\textbf{d}) Simulation of the charge induced on the qubit island from a unit point charge at various locations in the Si substrate.}
\label{fig:fig1}
\end{figure}

\begin{figure*}[ht!]
\includegraphics[width=\columnwidth]{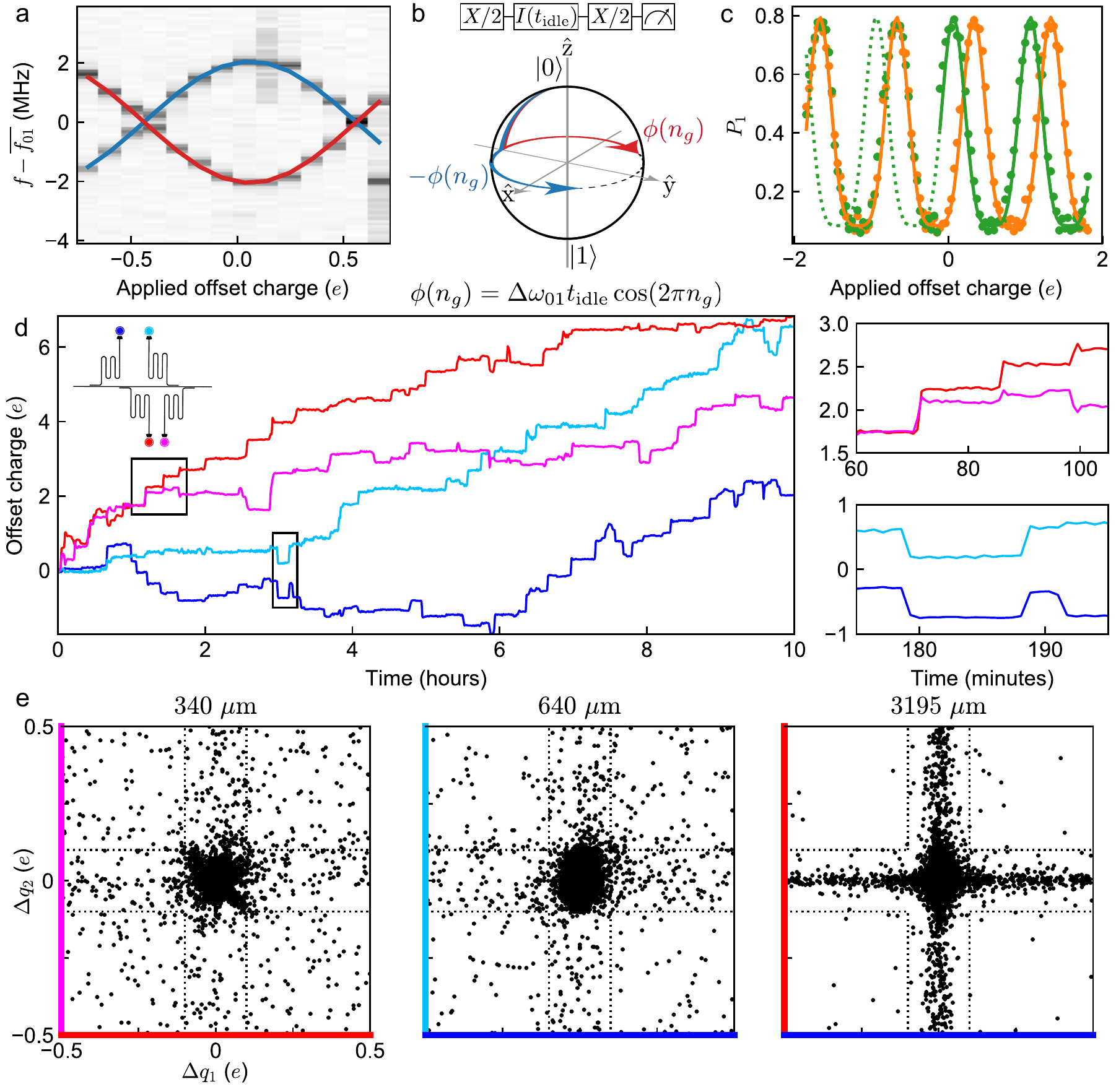}
\caption{ \textbf{Characterization of correlated charge fluctuations.} (\textbf{a}) Qubit spectroscopy \textit{versus} applied offset charge showing the two quasiparticle parity bands; a discrete jump in offset charge can be seen in the rightmost column of data.  (\textbf{b}) Ramsey sequence used to detect offset charge $n_g \equiv \Delta q/2e$, and trajectory of the qubit state vector for the two values of quasiparticle  parity. (\textbf{c}) Two sequential scans of Ramsey amplitude \textit{versus} offset charge; points are data and solid traces are theoretical fits. In the first scan (orange points), the offset charge was constant throughout the acquisition, while in the second scan (green points) a discrete jump in offset charge occurred during the scan. (\textbf{d}) Time series of offset charge on the four qubits measured simultaneously over 10~hours.  Trace colors identify the locations of the four qubits, as shown in the figure inset. Panels to the right show detailed views of correlated offset charge jumps in qubit pairs. (\textbf{e}) Joint charge histograms measured on three qubit pairs; coloring of axes encodes the qubit location, and center-to-center separation is shown above the plots.}
\label{fig:fig2}
\end{figure*}

The sample geometry is shown in Fig.~\ref{fig:fig1}a-c. The $6.25 \times 6.25$~mm$^2$ chip incorporates four weakly charge-sensitive transmon qubits with a ratio of Josephson energy to single-electron charging energy $E_J/E_C = 24$; the detailed device parameters are presented in the Supplement. Each qubit comprises a circular superconducting Nb island with radius $r_i=$~70~$\mu$m embedded in a circular cavity with radius $r_o=$~90.5~$\mu$m in the superconducting Nb groundplane. Two of the qubits are positioned on one side of the chip with a center-to-center separation of 340~$\mu$m, while two of the qubits are on the opposite side of the chip with 640~$\mu$m center-to-center separation. The qubit pairs are separated by around 3~mm. Each qubit is coupled to a local readout resonator that is in turn coupled to a common feedline. 

For the purposes of interpreting the experiments described here, it is useful to consider the qubits as electrometers with a large effective sensing area for electric fields in the substrate. For our concentric qubit geometry, the sensing area for uniform electric fields is $\pi \epsilon r_i r_o $, where $\epsilon$ is the relative permittivity of the medium. In Fig.~\ref{fig:fig1}d we display the numerically computed offset charge on the qubit island induced by a unit point charge at various locations in the substrate; for an applied unit charge, the induced offset charge is simply the fraction of electric field lines that terminate on the qubit island.
 
In a first series of experiments, we perform simultaneous Ramsey tomography on the four qubits to generate time series of fluctuating offset charge. In Fig.~\ref{fig:fig2}a we show representative qubit spectroscopy and in Fig.~\ref{fig:fig2}b we show the experimental pulse sequence for the charge measurements \cite{Christensen19}. The Ramsey $X/2-Idle-X/2$ sequence maps precession frequency to occupation of the qubit $\ket{1}$ state irrespective of the quasiparticle parity of the qubit island. We perform a series of such experiments for different applied gate charge as shown in Fig.~\ref{fig:fig2}c; the phase of the resulting curve reveals the offset charge on the qubit island. Note that this approach only allows measurement of offset charge modulo the fundamental charge $e$; large discrete jumps in offset charge will be aliased to the interval from -0.5$e$ to +0.5$e$. 

In Fig. \ref{fig:fig2}d, we show a typical time series of offset charge measured on the four qubits simultaneously. The Ramsey-based charge measurement involves 3000 projections of the qubits across 10 applied gate charges, with a total cycle time of 44 seconds. Focusing on large discrete changes $\Delta q$ in offset charge in the range $0.1e<|\Delta q| \leq 0.5e$, we find a rate of charge jumps of $1.35(3)$ mHz averaged over the four qubits. The right panel shows the detailed structure of the charge traces for nearest-neighbor pairs measured at shorter timescales. We observe numerous simultaneous discrete jumps in the offset charge of neighboring qubits. In Fig. \ref{fig:fig2}e we show joint histograms of charge jumps measured in various qubit pairs. For all qubits, there is a Gaussian peak at the center of the distribution due to experimental uncertainty in the reconstructed offset charge. For the pairs separated by 340 and 640~$\mu$m, however, we find many simultaneous discrete changes in offset charge. Again focusing on large charge jumps in the range $0.1e<|\Delta q| \leq 0.5e$ and correcting for random coincidence, we find a correlation probability of $54(4)\%$ for the qubit pair separated by 340~$\mu$m and a correlation probability of $46(4)\%$ for the qubit pair separated by 640~$\mu$m (see Supplement). For qubits on opposite sides of the chip with separation of order 3~mm, the rate of simultaneous charge jumps is consistent with random coincidence.  

As mentioned above, the characteristic length $\sqrt{r_i r_o}$ sets the scale over which charge is sensed in the bulk substrate. The high degree of correlation in charge fluctuations sensed by qubits with 640~$\mu$m separation indicates charging events with a large spatial footprint. There are two obvious candidates for such events: absorption of cosmic-ray muons in the qubit substrate and absorption of $\gamma$-rays from background radioactivity \cite{Vepsalainen20}. These events deposit energy of order 100~keV in the qubit substrate, roughly ten orders of magnitude greater than the $\sim$10~$\mu$eV energy scale of the qubit states. In both cases, the absorption event liberates charge in the substrate; a significant fraction of the free charge diffuses over hundreds of microns, leading to a large spatial footprint for the charging event that can be sensed by multiple qubits.

We perform detailed numerical modeling of charge bursts induced by the absorption of cosmic rays and background radioactivity. We use the GEANT4 toolkit~\cite{AGOSTINELLI2003250,Allison_1610988,ALLISON2016186} to calculate the energy deposited in the Si substrate. A simplified model of the cryostat (including vacuum can, radiation shields, stage plates, etc.) is used to calculate the flux of muons and $\gamma$-rays at the chip (see Supplement). The angular and energy distribution of simulated muons reproduces measurements of cosmic ray muons at sea level~\cite{shukla2018energy}, and the photons from background radioactivity are generated isotropically according to the energy distribution measured at Laboratori Nazionali del Gran Sasso (LNGS), which matches the distribution measured in the qubit lab in Madison (see Supplement).

\begin{figure}[t!]
\includegraphics[width=\columnwidth]{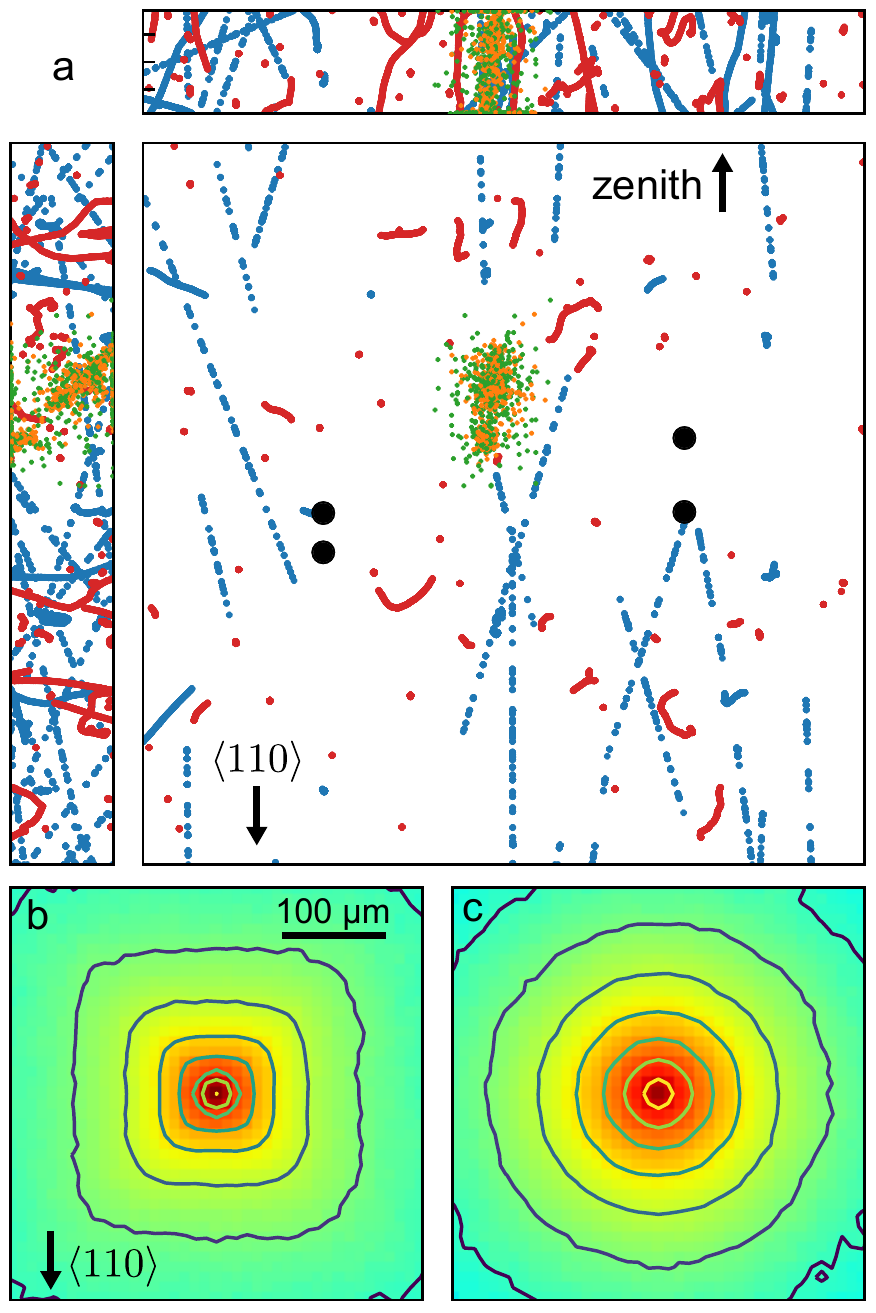}
\caption{ \textbf{Modeling of muon and $\gamma$-ray impacts.} (\textbf{a}) Top and side views of muon (blue, 30 events) and $\gamma$-ray (red, 60 events) tracks for a collection of simulated impact events in the 6.25$\times$6.25~mm$^2$ chip.  For a single muon track, a sample distribution of generated electrons (green) and holes (orange) is shown.  Qubit locations are indicated by black circles. The crystal orientation of the Si substrate is $\expect{001}$; the crystal $\expect{110}$ direction is as indicated. The chip is oriented within the cryostat as indicated (orientation is relevant for the simulation of cosmic ray muons, which predominantly arrive from the vertical direction). Electron (\textbf{b}) and hole (\textbf{c}) probability distributions used to simulate induced offset charge. Electrons tend to travel along the crystal valleys, while the distribution of holes is spherically symmetric. 
}
\label{fig:fig3}
\end{figure}

Each energy deposit liberates one electron-hole pair per 3.6~eV of energy transferred to the substrate~\cite{ramanathan2020ionization}. The subsequent diffusion of charge is modeled using G4CMP~\cite{brandt2014semiconductor,g4cmp}. This charge transport simulation takes into account anisotropy in the electron band structure, which leads to a separation of the positive and negative charge liberated by the burst event, as demonstrated in ref.~\citenum{SiChargeTransport}. The characteristic trapping length $\lambda_{\rm trap}$ is taken to be energy- and species-independent; $\lambda_{\rm trap}$ and the charge production efficiency $f_q$ are tuned to match the experimentally measured charge histograms (see Supplement for details). We find for $\lambda_{\rm trap} = 300$~$\mu$m  and  $f_q=0.2$ that the simulated single- and two-qubit charge histograms are in good qualitative agreement with the measured histograms and provide a reasonable quantitative match with the correlation probabilities and charge asymmetries extracted from the data. The trapping length $\lambda_{\rm trap}$ is a critical materials parameter that determines the electrostatic coupling of particle impact events to nearby qubits. Based on our analysis, we conclude that the contribution of $\gamma$-rays to the measured rate of charge bursts is around 40 times that of cosmic ray muons (see Supplement). We infer a rate of $\gamma$ impacts on the 6.25$\times$6.25~mm$^2$ chip of 19.8(5)~mHz.

\begin{figure}[t]
\includegraphics[width=\columnwidth]{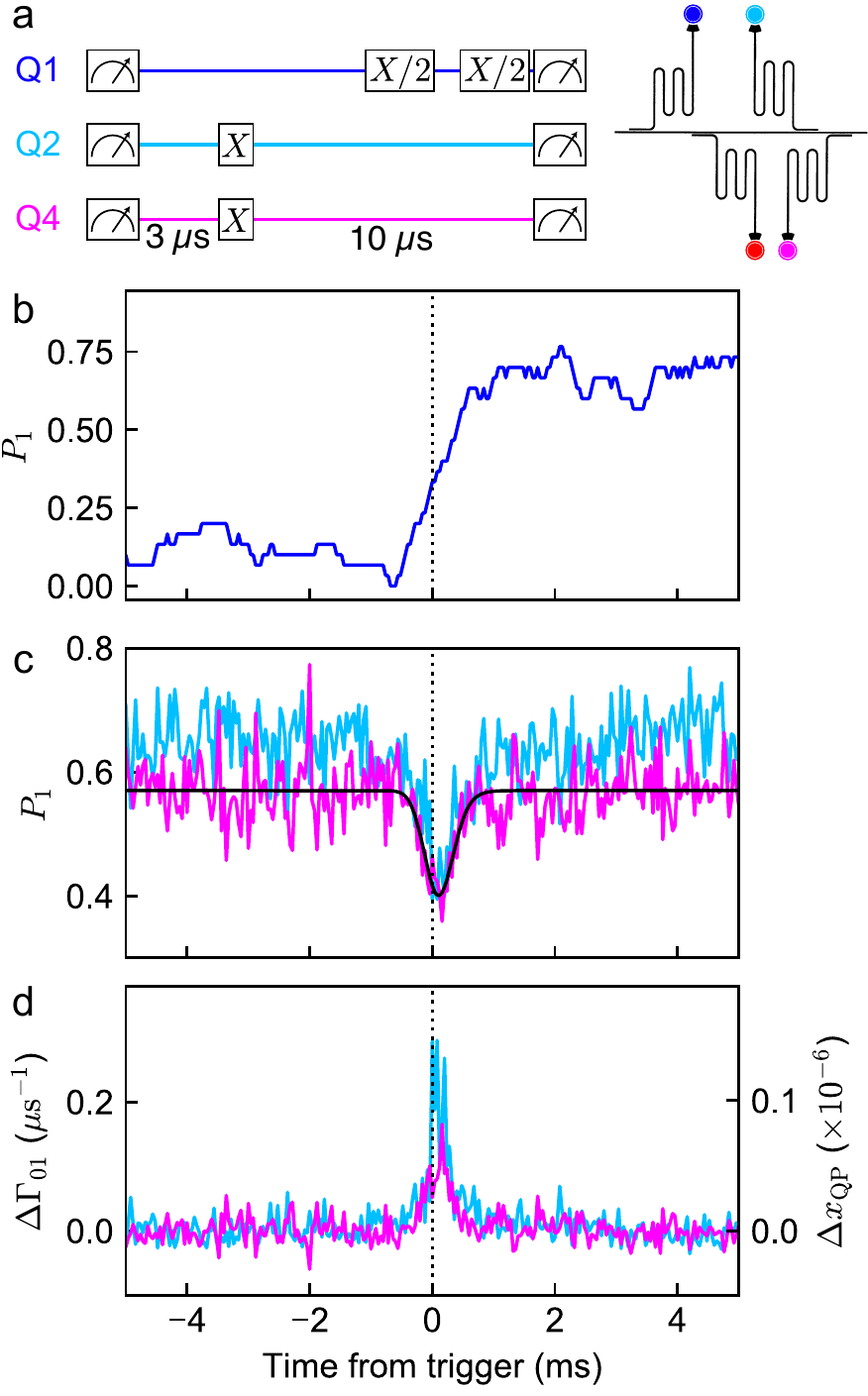}
\caption{ \textbf{Characterization of correlated relaxation errors.} (\textbf{a}) Experimental pulse sequence.  Qubit 1 (Q1) acts as a charge trigger, while qubits 2 and 4 (Q2, Q4) act as local probes of $T_1$.  
(\textbf{b}) Representative trace of the Ramsey amplitude of Q1 during a burst event; qubit occupation is averaged over 30 single-shot measurements. (\textbf{c}) Average single-shot occupation for Q2 (blue, 340~$\mu$m from Q1; 142 events) and Q4 (pink, 3~mm from Q1; 121 events) \textit{versus} time with respect to a detected charge burst.  Black trace is a fit to the data from Q4, yielding a recovery timescale 130$\pm$40~$\mu$s. (\textbf{d}) Average change $\Delta \Gamma_{01}$ in qubit relaxation rate and average change $\Delta x_{\rm QP}$ in reduced quasiparticle density calculated from the data in (c).}
\label{fig:fig4}
\end{figure}

Ultimately, the energy released by particle absorption will be transferred to the phonon reservoir in the qubit susbtrate. Phonons will rapidly scatter to the gap edge of the Nb groundplane by breaking Cooper pairs; nonequilibrium quasiparticles in the vicinity of the Al junction leads are expected to become trapped and suppress qubit relaxation time $T_1$ \cite{Martinis09,Lenander11,Wenner13,Wang14,Serniak18}. In a separate series of experiments, we use one qubit as a trigger for charge bursts while additional qubits are used as local probes of $T_1$. Fig. \ref{fig:fig4}a shows the pulse sequence for the experiment. On qubit 1 (Q1) we perform the same charge Ramsey sequence as in Fig. \ref{fig:fig2}, while on qubits 2 and 4 (Q2, Q4) we perform a stripped-down inversion recovery experiment consisting of a premeasurement to initialize the qubit, an $X$-gate, a fixed idle time of 10~$\mu$s, and a second measurement. The sequence is repeated continuously with a cycle time of 40~$\mu$s.  

We identify burst events when there is a large discrete change in the running average of the Ramsey amplitude measured on Q1, allowing us to align and average traces from the probe qubits.  In the absence of burst events, the inversion recovery sequence yields average occupations of the qubit $\ket{1}$ state that are consistent with the separately measured qubit $T_1$ times. When a charge burst is detected in Q1, however, we find a clear suppression in the $\ket{1}$ occupation of Q2 and Q4.  Fitting this dropout with an exponential recovery convolved with a Gaussian to account for timing uncertainty in the trigger, we find a recovery timescale of $130\pm40~\mu$s. We conclude that the same process that gives rise to discrete jumps in offset charge also leads to correlated suppression of qubit $T_1$ time over length scales in excess of 3~mm. In general, quasiparticles that trap in the junction leads in the immediate aftermath of a particle impact event will induce both upward and downward qubit transitions\cite{Catelani11b} that will be correlated across the qubit array.

The recovery timescale for quasiparticle poisoning can be understood in the following way. 
Phonons propagate diffusively to the boundary of the chip in a time $x_0^2/c_sz_0$, where $x_0$~=~6.25~mm is the lateral dimension of the chip, $c_s=6 \times 10^3$~m/s is the sound speed in the Si substrate, and $z_0=$~375~$\mu$m is the chip thickness. The chip is thermally anchored at four corners, with a fraction $\beta=0.2$ of the chip perimeter acoustically coupled to the chip enclosure. We therefore find a characteristic dwell time for athermal phonons in the substrate of order $x_0^2/\beta c_sz_0 \sim$~100~$\mu$s, in qualitative agreement with the measured recovery time. 

We briefly discuss the implications of these results for the realization of fault-tolerant superconducting qubit arrays in the surface code; for a detailed discussion, see Supplement. We define correlation degree $m$ as the number of qubits in a line to which an error couples. It can be shown that the fault-tolerant threshold $p_m$ for correlated errors of degree $m$ is given by $p_m \approx p^m$, where $p$ is the fault-tolerant threshold for uncorrelated errors \cite{Faoro20}. The relaxation and bit-flip errors associated with phonon-mediated quasiparticle poisoning are particularly damaging, as they couple to all qubits in a mm-scale chip. We identify two additional correlated error mechanisms: correlated phase-flip errors due to exponentially small (but  nonzero)  frequency  shifts  induced  by  correlated charge noise, and  correlated  bit-flip  errors  induced  by  the sudden charge transient associated with particle impact. Even for a nominally charge-insensitive qubit such as the transmon with $E_C/h$~=~250~MHz and $E_J/E_C = 50$, we find that the rate of correlated phase-flip errors is significant, with 0.9\% (3.8\%) of $\gamma$-ray (muon) impacts giving rise to correlated phase-flip errors above the $10^{-4}$ level in qubit pairs separated by 640~$\mu$m, and with 7.2\% of muon impacts giving rise to correlated phase-flip errors above the $10^{-6}$ level in qubit pairs separated by 3~mm. In general, the exponential sensitivity of the qubit array to correlated errors represents a serious design constraint: for a given error mechanism with fixed spatial footprint, the need to protect against correlated errors will dictate how closely spaced the qubits can be. 

A clear understanding of the underlying physics of particle impact events in the qubit substrate will allow the development of mitigation strategies to suppress or even eliminate correlated errors. We discuss several possible approaches below.

First, one can operate the quantum processor in a clean environment that provides shielding against cosmic ray muons and background $\gamma$-rays. Such measures are routinely taken in ultrasensitive searches for rare events, such as neutrinoless double beta decay~\cite{Dell_Oro_2016,Poda_2017} or dark matter interactions~\cite{baudis_2018,Pirro2017}. Underground sites enable the reduction of cosmic-ray muon flux to negligible levels \cite{Sz_cs_2019,Aglietta1998,Jillings_2016}. 
Similarly, the cryostat can be shielded in massive lead and copper structures to absorb $\gamma$-rays. A few centimeters of lead shielding guarantees a suppression of the $\gamma$ flux by 1-2 orders of magnitude.
Finally, the materials used to construct the device and its enclosure can be selected to be radio-pure and processed through electrochemical treatments that remove surface contamination \cite{Alessandria_2013,Aprile_2011,BUSTO200235,Ilias_DB,cardani2020}. 


Second, one could reduce the sensitivity of the qubit to the burst events. Reduction of the size of the qubit island and reduction of the gap from the island to ground will limit the sensitivity of the qubit to electric fields in the substrate. It is important to note that the near-continuous groundplane in the geometry studied here provides excellent electrostatic screening against charge in the bulk. We anticipate that a multiqubit architecture that lacks a groundplane will be much more susceptible to correlated phase-flip errors induced by charge bursts. 

In order to combat quasiparticle-induced $T_1$ suppression, mitigation strategies could be adopted to prevent the direct diffusion of quasiparticles, for example involving superconducting bandgap engineering \cite{Aumentado04} or normal-metal quasiparticle traps \cite{Patel17,Martinis20}. Finally, steps could be taken to promote the relaxation of high-energy phonons below the gap edge and to enhance the rate of removal of phonons from the qubit substrate~\cite{beckman}. 
Modest improvements in the acoustic anchoring of the substrate could accelerate recovery of the chip following particle absorption, minimizing correlated relaxation errors due to quasiparticles. 


We acknowledge stimulating discussions with R. Barends, I. M. Pop, and J. M. Martinis. We thank S.~Pirro for helpful discussions and for sharing the results of his measurements of environmental radioactivity. We thank J. Engle for assistance with the calibration of the NaI scintillation detector used to characterize background radioactivity in the lab in Madison, and we thank A. Riswadkar for support with device fabrication. Work at UW-Madison was supported by the U.S. Department of Energy (DOE), Office of Science, Basic Energy Sciences (BES) under Award \#DE-SC0020313. Parts of this document were prepared using the resources of the Fermi National Accelerator Laboratory (Fermilab), a U.S. Department of Energy, Office of Science, HEP User Facility. Fermilab is managed by Fermi Research Alliance, LLC (FRA), acting under Contract No. DE-AC02-07CH11359. Contributions from J. L DuBois were performed under the auspices of the U.S. Department of Energy by Lawrence Livermore National Laboratory under Contract DE-AC52-07NA27344. The authors acknowledge use of facilities and instrumentation at the UW-Madison Wisconsin Centers for Nanoscale Technology, partially supported by the NSF through the University of Wisconsin Materials Research Science and Engineering Center (DMR-1720415). We acknowledge IARPA and Lincoln Laboratory for providing the TWPA used in some of these experiments.

\bibliographystyle{naturemag}
\bibliography{ChargeNoise_BibTeX,RFM_refs_Feb20}

\begin{thebibliography}{10}
\expandafter\ifx\csname url\endcsname\relax
  \def\url#1{\texttt{#1}}\fi
\expandafter\ifx\csname urlprefix\endcsname\relax\def\urlprefix{URL }\fi
\providecommand{\bibinfo}[2]{#2}
\providecommand{\eprint}[2][]{\url{#2}}

\bibitem{Fowler12}
\bibinfo{author}{Fowler, A.~G.}, \bibinfo{author}{Mariantoni, M.},
  \bibinfo{author}{Martinis, J.~M.} \& \bibinfo{author}{Cleland, A.~N.}
\newblock \bibinfo{title}{Surface codes: Towards practical large-scale quantum
  computation}.
\newblock \emph{\bibinfo{journal}{Phys. Rev. A}} \textbf{\bibinfo{volume}{86}},
  \bibinfo{pages}{032324} (\bibinfo{year}{2012}).

\bibitem{Barends14}
\bibinfo{author}{Barends, R.} \emph{et~al.}
\newblock \bibinfo{title}{Superconducting quantum circuits at the surface code
  threshold for fault tolerance}.
\newblock \emph{\bibinfo{journal}{Nature}} \textbf{\bibinfo{volume}{508}},
  \bibinfo{pages}{500--503} (\bibinfo{year}{2014}).

\bibitem{Sheldon2016}
\bibinfo{author}{Sheldon, S.} \emph{et~al.}
\newblock \bibinfo{title}{Characterizing errors on qubit operations via
  iterative randomized benchmarking}.
\newblock \emph{\bibinfo{journal}{Phys. Rev. A}} \textbf{\bibinfo{volume}{93}},
  \bibinfo{pages}{012301} (\bibinfo{year}{2016}).

\bibitem{Jeffrey14}
\bibinfo{author}{Jeffrey, E.} \emph{et~al.}
\newblock \bibinfo{title}{Fast accurate state measurement with superconducting
  qubits}.
\newblock \emph{\bibinfo{journal}{Phys. Rev. Lett.}}
  \textbf{\bibinfo{volume}{112}}, \bibinfo{pages}{190504}
  (\bibinfo{year}{2014}).

\bibitem{Walter17}
\bibinfo{author}{Walter, T.} \emph{et~al.}
\newblock \bibinfo{title}{Rapid high-fidelity single-shot dispersive readout of
  superconducting qubits}.
\newblock \emph{\bibinfo{journal}{Phys. Rev. Applied}}
  \textbf{\bibinfo{volume}{7}}, \bibinfo{pages}{054020} (\bibinfo{year}{2017}).

\bibitem{Opremcak20}
\bibinfo{author}{Opremcak, A.} \emph{et~al.}
\newblock \bibinfo{title}{High-fidelity measurement of a superconducting qubit
  using an on-chip microwave photon counter.} \eprint{arXiv:2008.02346}
  (\bibinfo{year}{2020}).

\bibitem{Faoro20}
\bibinfo{author}{Faoro, L.} \& \bibinfo{author}{Ioffe, L.~B.}
\newblock \bibinfo{title}{Critical behavior of surface code statistical
  models}.
\newblock \emph{\bibinfo{journal}{In preparation}}  (\bibinfo{year}{2020}).

\bibitem{Christensen19}
\bibinfo{author}{Christensen, B.~G.} \emph{et~al.}
\newblock \bibinfo{title}{Anomalous charge noise in superconducting qubits}.
\newblock \emph{\bibinfo{journal}{Phys. Rev. B}}
  \textbf{\bibinfo{volume}{100}}, \bibinfo{pages}{140503}
  (\bibinfo{year}{2019}).

\bibitem{Vepsalainen20}
\bibinfo{author}{Vepsäläinen, A.~P.} \emph{et~al.}
\newblock \bibinfo{title}{Impact of ionizing radiation on superconducting qubit
  coherence}.
\newblock \emph{\bibinfo{journal}{Nature}} \textbf{\bibinfo{volume}{584}},
  \bibinfo{pages}{551–556} (\bibinfo{year}{2020}).

\bibitem{AGOSTINELLI2003250}
\bibinfo{author}{Agostinelli, S.} \emph{et~al.}
\newblock \bibinfo{title}{{GEANT}4—a simulation toolkit}.
\newblock \emph{\bibinfo{journal}{Nucl. Instrum. Methods. Phys. Res. A}}
  \textbf{\bibinfo{volume}{506}}, \bibinfo{pages}{250--303}
  (\bibinfo{year}{2003}).

\bibitem{Allison_1610988}
\bibinfo{author}{{Allison}, J.} \emph{et~al.}
\newblock \bibinfo{title}{{GEANT}4 developments and applications}.
\newblock \emph{\bibinfo{journal}{IEEE Trans. Nucl. Sci.}}
  \textbf{\bibinfo{volume}{53}}, \bibinfo{pages}{270--278}
  (\bibinfo{year}{2006}).

\bibitem{ALLISON2016186}
\bibinfo{author}{Allison, J.} \emph{et~al.}
\newblock \bibinfo{title}{Recent developments in {GEANT}4}.
\newblock \emph{\bibinfo{journal}{Nucl. Instrum. Methods. Phys. Res. A}}
  \textbf{\bibinfo{volume}{835}}, \bibinfo{pages}{186--225}
  (\bibinfo{year}{2016}).

\bibitem{shukla2018energy}
\bibinfo{author}{Shukla, P.} \& \bibinfo{author}{Sankrith, S.}
\newblock \bibinfo{title}{Energy and angular distributions of atmospheric muons
  at the {E}arth.} \eprint{arXiv:1606.06907} (\bibinfo{year}{2018}).

\bibitem{ramanathan2020ionization}
\bibinfo{author}{Ramanathan, K.} \& \bibinfo{author}{Kurinsky, N.~A.}
\newblock \bibinfo{title}{Ionization yield in silicon for e{V}-scale
  electron-recoil processes.} \eprint{arXiv:2004.10709} (\bibinfo{year}{2020}).

\bibitem{brandt2014semiconductor}
\bibinfo{author}{Brandt, D.} \emph{et~al.}
\newblock \bibinfo{title}{Semiconductor phonon and charge transport {M}onte
  {C}arlo simulation using {GEANT}4.} \eprint{arXiv:1403.4984}
  (\bibinfo{year}{2014}).

\bibitem{g4cmp}
\bibinfo{author}{Kelsey, M.}, \bibinfo{author}{Agnese, R.},
  \bibinfo{author}{Brandt, D.} \& \bibinfo{author}{Redl, P.}
\newblock \bibinfo{title}{{G4CMP}: {GEANT}4 add-on framework for phonon and
  charge-carrier physics}.
\newblock \bibinfo{howpublished}{https://github.com/kelseymh/G4CMP}
  (\bibinfo{year}{2020}).

\bibitem{SiChargeTransport}
\bibinfo{author}{Moffatt, R.~A.} \emph{et~al.}
\newblock \bibinfo{title}{Spatial imaging of charge transport in silicon at low
  temperature}.
\newblock \emph{\bibinfo{journal}{App. Phys. Lett.}}
  \textbf{\bibinfo{volume}{114}}, \bibinfo{pages}{032104}
  (\bibinfo{year}{2019}).

\bibitem{Martinis09}
\bibinfo{author}{Martinis, J.~M.}, \bibinfo{author}{Ansmann, M.} \&
  \bibinfo{author}{Aumentado, J.}
\newblock \bibinfo{title}{Energy decay in superconducting {J}osephson-junction
  qubits from nonequilibrium quasiparticle excitations}.
\newblock \emph{\bibinfo{journal}{Phys. Rev. Lett.}}
  \textbf{\bibinfo{volume}{103}}, \bibinfo{pages}{097002}
  (\bibinfo{year}{2009}).

\bibitem{Lenander11}
\bibinfo{author}{Lenander, M.} \emph{et~al.}
\newblock \bibinfo{title}{Measurement of energy decay in superconducting qubits
  from nonequilibrium quasiparticles}.
\newblock \emph{\bibinfo{journal}{Phys. Rev. B}} \textbf{\bibinfo{volume}{84}},
  \bibinfo{pages}{024501} (\bibinfo{year}{2011}).

\bibitem{Wenner13}
\bibinfo{author}{Wenner, J.} \emph{et~al.}
\newblock \bibinfo{title}{Excitation of superconducting qubits from hot
  nonequilibrium quasiparticles}.
\newblock \emph{\bibinfo{journal}{Phys. Rev. Lett.}}
  \textbf{\bibinfo{volume}{110}}, \bibinfo{pages}{150502}
  (\bibinfo{year}{2013}).

\bibitem{Wang14}
\bibinfo{author}{Wang, C.} \emph{et~al.}
\newblock \bibinfo{title}{Measurement and control of quasiparticle dynamics in
  a superconducting qubit}.
\newblock \emph{\bibinfo{journal}{Nat. Commun.}} \textbf{\bibinfo{volume}{5}},
  \bibinfo{pages}{5836} (\bibinfo{year}{2014}).

\bibitem{Serniak18}
\bibinfo{author}{Serniak, K.} \emph{et~al.}
\newblock \bibinfo{title}{Hot nonequilibrium quasiparticles in transmon
  qubits}.
\newblock \emph{\bibinfo{journal}{Phys. Rev. Lett.}}
  \textbf{\bibinfo{volume}{121}}, \bibinfo{pages}{157701}
  (\bibinfo{year}{2018}).

\bibitem{Catelani11b}
\bibinfo{author}{Catelani, G.}, \bibinfo{author}{Schoelkopf, R.~J.},
  \bibinfo{author}{Devoret, M.~H.} \& \bibinfo{author}{Glazman, L.~I.}
\newblock \bibinfo{title}{Relaxation and frequency shifts induced by
  quasiparticles in superconducting qubits}.
\newblock \emph{\bibinfo{journal}{Phys. Rev. B}} \textbf{\bibinfo{volume}{84}}
  (\bibinfo{year}{2011}).

\bibitem{Dell_Oro_2016}
\bibinfo{author}{Dell’Oro, S.}, \bibinfo{author}{Marcocci, S.},
  \bibinfo{author}{Viel, M.} \& \bibinfo{author}{Vissani, F.}
\newblock \bibinfo{title}{Neutrinoless double beta decay: 2015 review}.
\newblock \emph{\bibinfo{journal}{Adv. High Energy Phys.}}
  \textbf{\bibinfo{volume}{2016}}, \bibinfo{pages}{1–37}
  (\bibinfo{year}{2016}).

\bibitem{Poda_2017}
\bibinfo{author}{Poda, D.} \& \bibinfo{author}{Giuliani, A.}
\newblock \bibinfo{title}{Low background techniques in bolometers for
  double-beta decay search}.
\newblock \emph{\bibinfo{journal}{Int. J. Mod. Phys. A}}
  \textbf{\bibinfo{volume}{32}}, \bibinfo{pages}{1743012}
  (\bibinfo{year}{2017}).

\bibitem{baudis_2018}
\bibinfo{author}{Baudis, L.}
\newblock \bibinfo{title}{The search for dark matter}.
\newblock \emph{\bibinfo{journal}{Eur. Rev.}} \textbf{\bibinfo{volume}{26}},
  \bibinfo{pages}{70–81} (\bibinfo{year}{2018}).

\bibitem{Pirro2017}
\bibinfo{author}{Pirro, S.} \& \bibinfo{author}{Mauskopf, P.}
\newblock \bibinfo{title}{Advances in bolometer technology for fundamental
  physics}.
\newblock \emph{\bibinfo{journal}{Annu. Rev. Nucl. Part. Sci.}}
  \textbf{\bibinfo{volume}{67}}, \bibinfo{pages}{161--181}
  (\bibinfo{year}{2017}).

\bibitem{Sz_cs_2019}
\bibinfo{author}{Szücs, T.} \emph{et~al.}
\newblock \bibinfo{title}{Background in $\gamma$-ray detectors and carbon beam
  tests in the {F}elsenkeller shallow-underground accelerator laboratory}.
\newblock \emph{\bibinfo{journal}{Eur. Phys. J. A}}
  \textbf{\bibinfo{volume}{55}}, \bibinfo{pages}{174} (\bibinfo{year}{2019}).

\bibitem{Aglietta1998}
\bibinfo{author}{Aglietta, M.} \emph{et~al.}
\newblock \bibinfo{title}{Muon ``depth-intensity'' relation measured by the lvd
  underground experiment and cosmic-ray muon spectrum at sea level}.
\newblock \emph{\bibinfo{journal}{Phys. Rev. D}} \textbf{\bibinfo{volume}{58}},
  \bibinfo{pages}{092005} (\bibinfo{year}{1998}).

\bibitem{Jillings_2016}
\bibinfo{author}{Jillings, C.}
\newblock \bibinfo{title}{The {SNOLAB} science program}.
\newblock \emph{\bibinfo{journal}{J. Phys. Conf. Ser.}}
  \textbf{\bibinfo{volume}{718}}, \bibinfo{pages}{062028}
  (\bibinfo{year}{2016}).

\bibitem{Alessandria_2013}
\bibinfo{author}{Alessandria, F.} \emph{et~al.}
\newblock \bibinfo{title}{Validation of techniques to mitigate copper surface
  contamination in {CUORE}}.
\newblock \emph{\bibinfo{journal}{Astropart. Phys.}}
  \textbf{\bibinfo{volume}{45}}, \bibinfo{pages}{13–22}
  (\bibinfo{year}{2013}).

\bibitem{Aprile_2011}
\bibinfo{author}{Aprile, E.} \emph{et~al.}
\newblock \bibinfo{title}{Material screening and selection for {XENON100}}.
\newblock \emph{\bibinfo{journal}{Astropart. Phys.}}
  \textbf{\bibinfo{volume}{35}}, \bibinfo{pages}{43–49}
  (\bibinfo{year}{2011}).

\bibitem{BUSTO200235}
\bibinfo{author}{Busto, J.}, \bibinfo{author}{Gonin, Y.},
  \bibinfo{author}{Hubert, F.}, \bibinfo{author}{Hubert, P.} \&
  \bibinfo{author}{Vuilleumier, J.-M.}
\newblock \bibinfo{title}{Radioactivity measurements of a large number of
  adhesives}.
\newblock \emph{\bibinfo{journal}{Nucl. Instrum. Methods. Phys. Res. A}}
  \textbf{\bibinfo{volume}{492}}, \bibinfo{pages}{35--42}
  (\bibinfo{year}{2002}).

\bibitem{Ilias_DB}
\bibinfo{title}{{ILIAS Database}}.
\newblock \bibinfo{howpublished}{\url{http://radiopurity.in2p3.fr}}.

\bibitem{cardani2020}
\bibinfo{author}{Cardani, L.} \emph{et~al.}
\newblock \bibinfo{title}{Reducing the impact of radioactivity on quantum
  circuits in a deep-underground facility.} \eprint{arXiv:2005.02286}
  (\bibinfo{year}{2020}).

\bibitem{Aumentado04}
\bibinfo{author}{Aumentado, J.}, \bibinfo{author}{Keller, M.~W.},
  \bibinfo{author}{Martinis, J.~M.} \& \bibinfo{author}{Devoret, M.~H.}
\newblock \bibinfo{title}{Nonequilibrium quasiparticles and $2e$ periodicity in
  single-{C}ooper-pair transistors}.
\newblock \emph{\bibinfo{journal}{Phys. Rev. Lett.}}
  \textbf{\bibinfo{volume}{92}}, \bibinfo{pages}{066802}
  (\bibinfo{year}{2004}).

\bibitem{Patel17}
\bibinfo{author}{Patel, U.}, \bibinfo{author}{Pechenezhskiy, I.~V.},
  \bibinfo{author}{Plourde, B. L.~T.}, \bibinfo{author}{Vavilov, M.~G.} \&
  \bibinfo{author}{McDermott, R.}
\newblock \bibinfo{title}{Phonon-mediated quasiparticle poisoning of
  superconducting microwave resonators}.
\newblock \emph{\bibinfo{journal}{Phys. Rev. B}} \textbf{\bibinfo{volume}{96}},
  \bibinfo{pages}{220501(R)} (\bibinfo{year}{2017}).

\bibitem{Martinis20}
\bibinfo{author}{Martinis, J.~M.}
\newblock \bibinfo{title}{Saving superconducting quantum processors from qubit
  decay and correlated errors generated by gamma and cosmic rays.}
  \eprint{arXiv:2012.06137} (\bibinfo{year}{2020}).

\bibitem{beckman}
\bibinfo{author}{Beckman, S.~M.} \emph{et~al.}
\newblock \bibinfo{title}{{Development of cosmic ray mitigation techniques for
  the LiteBIRD space mission}}.
\newblock In \bibinfo{editor}{Zmuidzinas, J.} \& \bibinfo{editor}{Gao, J.-R.}
  (eds.) \emph{\bibinfo{booktitle}{Millimeter, Submillimeter, and Far-Infrared
  Detectors and Instrumentation for Astronomy IX}}, vol.
  \bibinfo{volume}{10708}. \bibinfo{organization}{International Society for
  Optics and Photonics} (\bibinfo{publisher}{SPIE}, \bibinfo{year}{2018}).

\bibitem{Dolan77}
\bibinfo{author}{Dolan, G.~J.}
\newblock \bibinfo{title}{Offset masks for lift-off photoprocessing}.
\newblock \emph{\bibinfo{journal}{Appl. Phys. Lett.}}
  \textbf{\bibinfo{volume}{31}}, \bibinfo{pages}{337--339}
  (\bibinfo{year}{1977}).

\bibitem{Jacoboni}
\bibinfo{author}{Jacoboni, C.} \& \bibinfo{author}{Reggiani, L.}
\newblock \bibinfo{title}{The {M}onte {C}arlo method for the solution of charge
  transport in semiconductors with applications to covalent materials}.
\newblock \emph{\bibinfo{journal}{Rev. Mod. Phys.}}
  \textbf{\bibinfo{volume}{55}}, \bibinfo{pages}{645--705}
  (\bibinfo{year}{1983}).

\bibitem{Sundqvist:2012wya}
\bibinfo{author}{Sundqvist, K.~M.}
\newblock \emph{\bibinfo{title}{{Carrier Transport and Related Effects in
  Detectors of the Cryogenic Dark Matter Search}}}.
\newblock Ph.D. thesis, \bibinfo{school}{UC-Berkeley} (\bibinfo{year}{2012}).

\bibitem{Moffatt:2016kok}
\bibinfo{author}{Moffatt, R.}
\newblock \emph{\bibinfo{title}{{Two-Dimensional Spatial Imaging of Charge
  Transport in Germanium Crystals at Cryogenic Temperatures}}}.
\newblock Ph.D. thesis, \bibinfo{school}{Stanford University}
  (\bibinfo{year}{2016}).

\bibitem{JKoch07}
\bibinfo{author}{Koch, J.} \emph{et~al.}
\newblock \bibinfo{title}{Charge-insensitive qubit design derived from the
  {C}ooper pair box}.
\newblock \emph{\bibinfo{journal}{Phys. Rev. A}} \textbf{\bibinfo{volume}{76}},
  \bibinfo{pages}{042319} (\bibinfo{year}{2007}).

\bibitem{Courbois1969}
\bibinfo{author}{Courbois, T.}, \bibinfo{author}{Van~Gelderen, L.} \&
  \bibinfo{author}{Leutz, H.}
\newblock \bibinfo{title}{Background, peak efficiency and dimensions of
  {NaI(Tl)}-crystals}.
\newblock \emph{\bibinfo{journal}{Nucl. Instrum. Methods}}
  \textbf{\bibinfo{volume}{69}}, \bibinfo{pages}{93--100}
  (\bibinfo{year}{1969}).

\end{thebibliography}

\vspace{12pt}

\newpage
\clearpage

\section*{SUPPLEMENT}
\beginsupplement

\subsection{Device Fabrication}
Our devices are fabricated in a single-layer process on high-resistivity ($>$10 k$\Omega$~cm) Si(001) wafers.
Following a hydrofluoric acid strip of the native SiO$_x$, we sputter a 90-nm film of Nb at a rate of 45~nm/min. We cap this film \textit{in situ} with a 20-nm layer of Al grown at a rate of 8 nm/min. The Nb deposition parameters are tuned to yield films with slight compressive stress.  We use an i-line projection aligner to define the qubit islands and all readout and control structures, and we etch the metal using a Cl$_2$/BCl$_3$ recipe in an inductively coupled plasma reactive ion etch tool.  The qubit junctions are realized using a standard Dolan bridge process \cite{Dolan77}. We pattern the MMA/PMMA stack with a 100-keV electron-beam writer. We shadow evaporate the Al-AlO$_{\rm x}$-Al stack in an electron-beam evaporation tool with base pressure $1\times10^{-8}$~Torr; prior to junction growth we perform an ion mill clean of the substrate to ensure good metallic contact to the base metal layer.

\subsection{Circuit Parameters}

In Table \ref{tab:qparams}, we list the designed and measured parameters of the devices used in these experiments.

\begin{table*}
\setlength{\tabcolsep}{10pt}
\renewcommand{\arraystretch}{1.1}
\begin{tabular}{ c c c c c c c }
  & Quantity & Q1 & Q2 & Q3 & Q4 & Units \\ \hline
 $g/2\pi$ & Designed qubit-resonator coupling & \multicolumn{4}{c}{27} & MHz \\
 $\kappa$ & Designed decay rate of readout resonators & \multicolumn{4}{c}{295} & $\mathrm{ns}^{-1}$ \\
 $f_{01}$ & Mean qubit transition frequency & 4.5641 & 4.4330 & 4.2939 & 4.3973 & GHz \\
 $f_r$ & Measured frequency of readout mode & 6.195 & 5.835 & 6.082 & 5.966 & GHz \\
 $2\Delta f_{01}$ & Measured pk-pk charge dispersion & 4.1 & 5.3 & 7.2 & 6.3 & MHz \\
 $\eta/2\pi$ & Measured qubit anharmonicity & 435 & 434 & 430 & -- & MHz 
\caption{\textbf{Parameters of devices used in the experiments.}}
\label{tab:qparams}
\end{tabular}
\end{table*}

\subsection{Measurement Setup}

\begin{figure*}
\includegraphics[width=\textwidth]{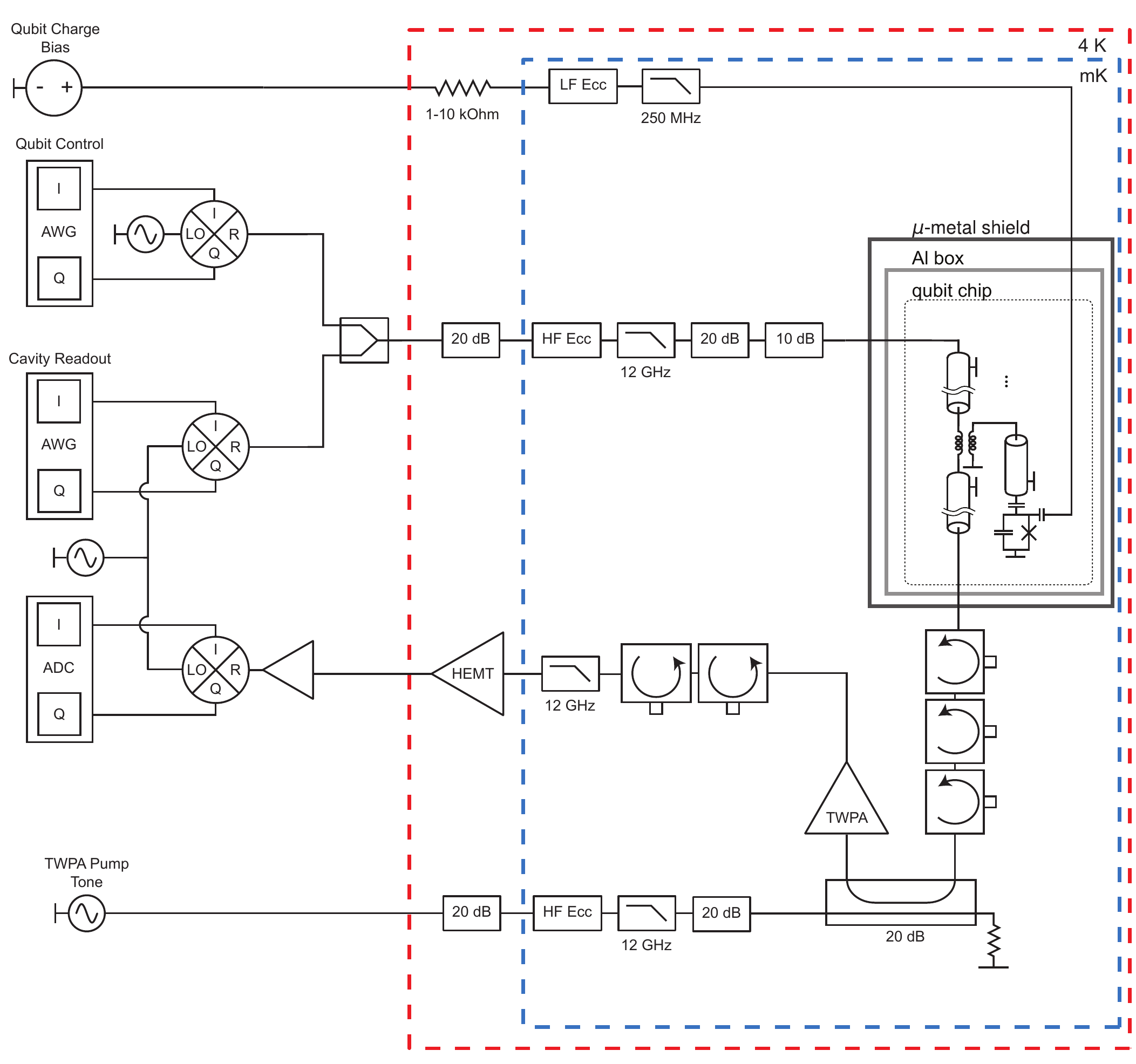}
\caption{\textbf{Wiring diagram of the experiments.}}
\label{fig:wiring_diagram} 
\end{figure*}

Our measurement setup is shown in Fig. \ref{fig:wiring_diagram}.  We use a standard microwave heterodyne setup involving single-sideband modulation of a local oscillator for both qubit and resonator drive.  Each microwave control line passes through multiple stages of filtering and attenuation at the 4~K and mK plates of the dilution refrigerator, including low-pass Eccosorb filters with a cutoff at 20~GHz.  Quasistatic charge bias lines pass through lower-frequency Eccosorb filters with a cutoff at 300~MHz.  For the measurements described in Fig. \ref{fig:fig4}, readout signals are preamplified by a traveling-wave parametric amplifier (TWPA) at the mK stage followed by a high electron mobility transistor (HEMT) amplifier at the 4~K stage.  The TWPA requires a separate microwave pump tone and can be switched in and out of the readout chain by a pair of microwave coaxial relays (not shown).   

\subsection{$T_1$ dropout}

We fit the $T_1$ dropout data of Fig. \ref{fig:fig4} with an exponential recovery convolved with a Gaussian to account for timing imprecision of our Ramsey-based charge trigger. We use the fit function  
\begin{equation}
    P_1(t) = \frac{1}{2} \exp\left(\frac{\sigma ^2-2 \tau t}{2 \tau ^2}\right) \text{erfc}\left(\frac{\sigma ^2-\tau t}{\sqrt{2} \sigma  \tau }\right)
\end{equation}
to extract the characteristic recovery time $\tau$, where $t$ is the elapsed time from the charge trigger and $\sigma$ is the standard deviation of the Gaussian. A fit to the data of Fig. \ref{fig:fig4}c yields a recovery time of $\tau=130\pm40$~$\mu$s and a timing jitter $\sigma=210\pm30$~$\mu$s.

We ascribe the average suppression in $P_1$ to an enhanced qubit relaxation rate $\Delta \Gamma_{01}$ due to quasiparticle poisoning. To convert the average change in relaxation rate to an increased density of nonequilibrium quasiparticles, we use the expression \cite{Wang14} 

\begin{equation}
    \Delta \Gamma_{01} = \frac{x_{\mathrm{QP}}}{\pi} \sqrt{\frac{2\Delta}{\hbar} \omega_{01}},
\end{equation}
where $x_{\rm QP}$ is the reduced density of quasiparticles (i.e., density of quasiparticles $n_{\rm QP}$ relative to the Cooper pair density $n_{\rm CP}$; for Al, $n_{\rm CP} = 4 \times 10^6$~$\mu$m$^{-3}$) and $2\Delta/e$=~380~$\mu$V is the superconducting gap of Al.

\subsection{Correlation Probabilities and Event Rates}

It is necessary to differentiate true correlated charging events in which a single charge burst is sensed by multiple qubits from apparent correlations that arise from our finite sampling rate. We label two qubits $A$ and $B$; we define $p^{\rm obs}_A$ ($p^{\rm obs}_B$) as the probability that a large discrete charge jump is registered in a given 44-second measurement cycle on qubit $A$ (qubit $B$), while $p^{\rm obs}_{AB}$ is the probability that we observe a jump in both qubits. The events that are seen by qubit $A$ have two contributions: some of these events are due to charge bursts that couple to both qubits (occurring with probability $p_{AB}$), while some are due to events that are seen only by qubit $A$ (occurring with probability $p_A$). We thus have 
\begin{equation} \label{pA}
    p^{\rm obs}_A = p_{AB} + p_A(1-p_{AB}),
\end{equation}
and similarly for $p^{\rm obs}_B$. In the same way, there are two contributions to $p^{\rm obs}_{AB}$: one contribution from burst events that couple to both qubits simultaneously (again, occurring with probability $p_{AB}$), and one contribution from random coincidence, where independent charging events, each seen by only one of the qubits, occur during the same measurement cycle. We have
\begin{equation} \label{pC}
    p^{\rm obs}_{AB} = p_{AB} + p_Ap_B(1-p_{AB}).
\end{equation}

Solving for the probability of true correlated events in terms of the observed probabilities, we find
\begin{equation}
    p_{AB} = \frac{p^{\rm obs}_{AB}-p^{\rm obs}_Ap^{\rm obs}_B}{1+p^{\rm obs}_{AB}-(p^{\rm obs}_A+p^{\rm obs}_B)}.
\end{equation}
In Table \ref{tab:corrprobs} we list the observed ($p_i^{\rm obs}, \, p_{ij}^{\rm obs}$) and inferred ($p_i, \, p_{ij}$) probabilities for discrete charge jumps per measurement cycle extracted from the dataset of Fig. \ref{fig:fig2}; here the indices $i,j$ specify the qubit or qubit pair. 

We define correlation probability $p_{AB}^{\rm corr}$ as the ratio of the probability of a true correlated event to the average of the probabilities of observed events in qubits $A$ and $B$:
\begin{equation}
    p_{AB}^{\rm corr} = \frac{2 p_{AB}}{p_A^{\rm obs} + p_B^{\rm obs}}.
\end{equation}
With this definition, $p_{AB}^{\rm corr}$ represents the probability that a large discrete offset charge jump seen by one qubit will also be seen by a neighboring qubit at a given separation. As we discuss below, comparison of simulated and measured values for $p_{AB}^{\rm corr}$ provides a means to pin down the charge trapping length scale and charge production efficiency associated with charge diffusion following an absorption event.  We obtain correlation probabilities of $0.54(4)$ for Q3-Q4 ($340~\mathrm{\mu m}$), $0.46(4)$ for Q1-Q2 ($640~\mathrm{\mu m}$), and $0.00(1)$ for Q1-Q3 ($3195~\mathrm{\mu m}$).

\begin{table}[b]
\setlength{\tabcolsep}{7pt}
\renewcommand{\arraystretch}{1.1}
\begin{tabular}{ c c c c c }
 &  &  &  & $\Gamma_i^\mathrm{obs},\Gamma_{ij}$ \\
 Separation & Qubit(s) & $p_i^\mathrm{obs},p_{ij}^\mathrm{obs}$ & $p_i,p_{ij}$ & (mHz) \\ 
 \hline
 - & Q1 & 0.055(2) & 0.030(3) & 1.27(5) \\
 - & Q2 & 0.061(3) & 0.035(4) & 1.38(6) \\
 - & Q3 & 0.060(3) & 0.029(3) & 1.38(6) \\
 - & Q4 & 0.060(3) & 0.029(4) & 1.38(6) \\
 \hline
 $340~\mathrm{\mu m}$ & Q3-Q4 & 0.033(2) & 0.033(2) & 0.74(6) \\
 $640~\mathrm{\mu m}$ & Q1-Q2 & 0.027(2) & 0.026(2) & 0.60(5) \\
 $3195~\mathrm{\mu m}$ & Q1-Q3 & 0.004(1) & 0.000(1) & 0.00(2)
\caption{\textbf{Probabilities and rates of impact for qubits and qubit pairs.}}
\label{tab:corrprobs}
\end{tabular}
\end{table}

It is instructive to connect the observed rate of discrete charge jumps in the range $0.1e < |\Delta q| \leq 0.5e$ to an absolute rate of particle impacts on the qubit chip. In Table \ref{tab:corrprobs} we show the rates $\Gamma_i^\mathrm{obs}$ of large discrete jumps observed on each qubit along with the inferred rates of correlated jumps $\Gamma_{ij}$ in the various qubit pairs.  We find a rate of discrete charge jumps of 1.35(3) mHz averaged across the four qubits. From the GEANT4 simulations, we expect an absolute rate of muon impacts on the chip of 0.5~mHz. Similarly, from the GEANT4 and charge transport simulations described below, we know that 16\% of these muon events will lead to an aliased offset charge jump above the threshold of 0.1$e$. We ascribe the remaining jump events to $\gamma$-ray absorption in the qubit substrate, with rate 1.27(3)~mHz. Again using the results of the GEANT4 and charge transport simulations, we can map this rate of charge bursts seen by the individual qubits to a rate of $\gamma$ impacts on the qubit chip. We find that the only 6\% of $\gamma$-ray absorptions lead to an aliased offset charge jump above 0.1$e$, implying the rate of $\gamma$ impacts on the 6.25 $\times$ 6.25 mm$^2$ qubit chip is 19.8(5)~mHz. Thus, if we we consider only the rate of impacts on the qubit chip, the contribution of environmental radioactivity is roughly a factor 40 larger than that of cosmic ray muons. However, as we show below, the muon impact events lead to stronger correlations across qubit devices.

\subsection{Modeling of Charge Bursts}
\label{sec:modeling}
\begin{figure}
\includegraphics[width=\columnwidth]{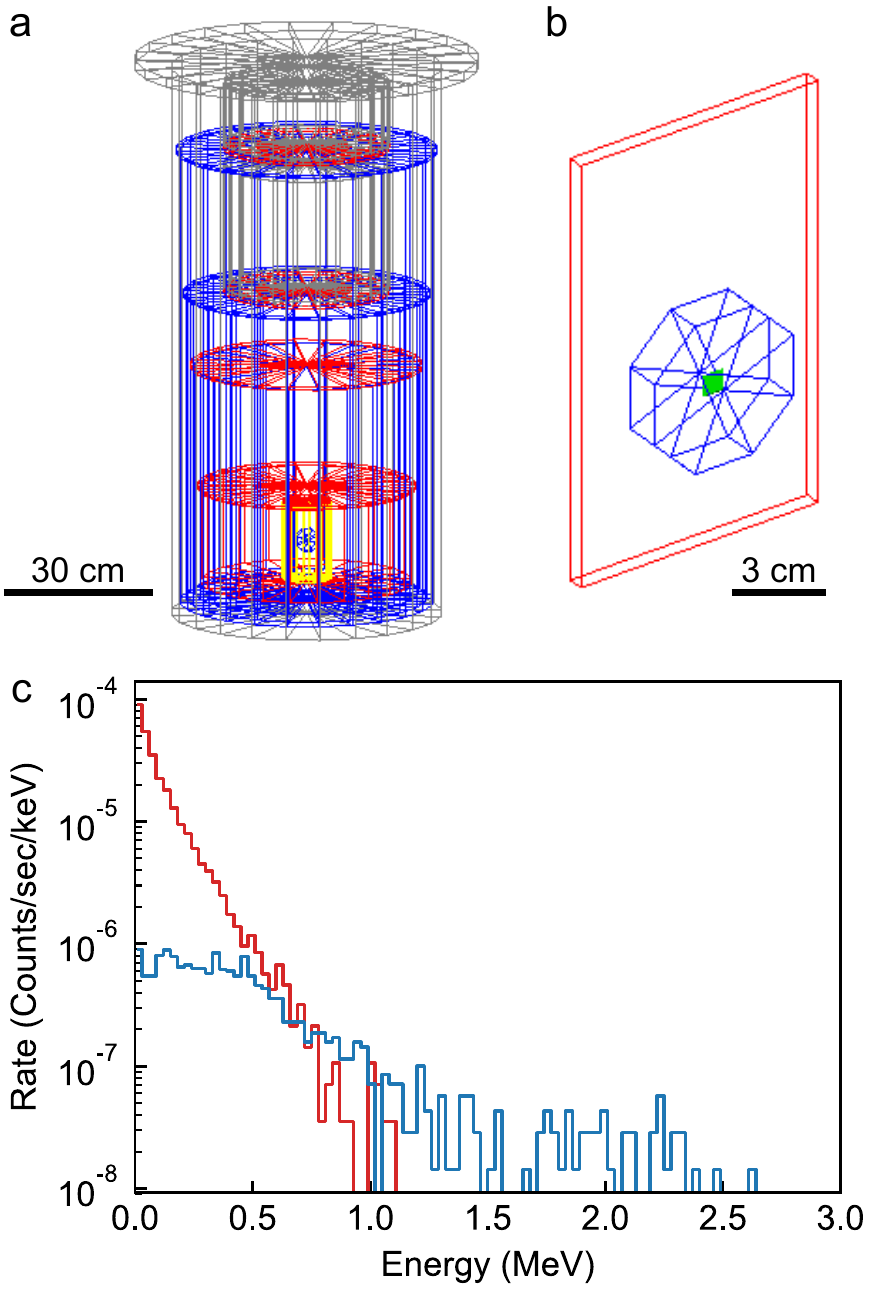}
\caption{ \textbf{GEANT4 modeling of the experiment.} (\textbf{a}) Model of the dilution refrigerator cryostat used in GEANT4 simulations of particle absorption events. (\textbf{b}) Model of the silicon chip, aluminum sample enclosure, and copper stage plate used in the simulations. The materials of the cryostat and chip enclosure include stainless steel (grey), copper (red), aluminum (blue), and cryoperm (yellow); the silicon chip is shown in green. (\textbf{c}) Energy deposited in the qubit substrate from environmental radioactivity (red) and cosmic ray muons (blue). The simulation of environmental radioactivity assumes a $\gamma$ flux of 2.8\,$\gamma$/cm$^2$/s. The $\gamma$-rays deposit an average energy of 100\,keV, while the average energy deposited by muons and their secondary $\gamma$-rays is 460\,keV.
}
\label{fig:geant}
\end{figure}

We use the GEANT4 software framework to model the absorption of energy in the qubit chip \cite{AGOSTINELLI2003250,Allison_1610988,ALLISON2016186}. We consider a simplified model of the dilution refrigerator cryostat, including the vacuum can, nested radiation shields, copper stage plates and sample stage, and aluminum sample enclosure; see Fig.~\ref{fig:geant}a-b. We focus on two contributions to the measured rate of charge bursts: $\gamma$-rays produced by environmental radioactivity and cosmic-ray muons.

Environmental radioactivity is commonly ascribed to $^{40}$K in addition to $^{232}$Th and $^{238}$U and their daughter nuclei. These contaminants are generally found throughout building materials, cryogenic infrastructure, and in the air itself. Their relative abundance and activity can vary from site to site. Nevertheless, the energy scale of interest is a few MeV, and the typical flux is of order 1~$\gamma$/cm$^2$/s. As input for the simulations, we use the spectrum of background radiation measured in Hall~C of the Laboratori Nazionali del Gran Sasso (LNGS) in Italy; as we discuss in Section~\ref{sec:bkg_measurement} below, this spectrum is comparable to that measured in the qubit laboratory in Madison apart from an overall reduction in the absolute rate by a factor 2.8, which does not impact the simulation results. 

We generate 10$^{10}$ $\gamma$-rays uniformly distributed on a cylindrical surface centred around the cryostat and record the track of each particle. Most $\gamma$-rays cross the setup without interacting, while a small fraction of $\gamma$-rays hit the qubit chip and produce an electron. Typical electron tracks produced by $\gamma$ interactions are shown in Fig.~\ref{fig:fig3} of the main text. The spectrum of energy deposited by these electrons is shown in Fig.~\ref{fig:geant}c. The spectrum extends to around 1~MeV with a mean energy deposit of 100~keV; the distribution of energy deposited in the substrate is insensitive to the detailed shape of the spectrum of incident $\gamma$-rays. 

Cosmic-ray muons provide a subdominant contribution to the measured rate of particle impacts. Muons release energy by ionization with a typical flux of 1\,$\mu$/cm$^2$/min. We generate 10$^8$ muons on a flat surface above the cryostat using the energy distribution and angular distribution reported in ref.~\citenum{shukla2018energy}. We record the track of each muon; like $\gamma$-rays, the vast majority of muons cross the setup without interacting, while a small fraction hit the substrate, producing a continuous track that deposits a mean energy of 460\,keV (see Fig.~\ref{fig:fig3} of the main text and Fig.~\ref{fig:geant}c). In some cases, the muons interact with the cryostat material, producing secondary $\gamma$-rays. The total rate of muon events in the substrate (primary + secondary $\gamma$ events) is 0.5\,mHz.

Using the tracks generated by GEANT4 as a starting point, we model the diffusion of electrons and holes in the qubit substrate. To construct the charge PDFs, we simulate $10^{8}$ electron-hole pairs originating uniformly along a line in the $z$-direction in the Si substrate (the direction normal to the chip surface). The substrate thickness and crystal orientation are chosen to match the parameters of the qubit chip under test; namely, we take a thickness of 375~$\mu$m, and we take the $\langle 0 0 1\rangle$ direction to be normal to the chip surface while the $\langle 1 1 0 \rangle$ and $\langle$1-10$\rangle$ directions are aligned with the chip edges. Each electron-hole pair is given 3.6~eV of initial total energy (2.6~eV of kinetic energy) and the momenta are randomized. In the case of electrons, randomization occurs in a spherically symmetric way before application of a Herring-Vogt transform to simulate the valley anisotropy\cite{Jacoboni}; the initial valley occupation is randomly chosen. The charges are then propagated and allowed to emit phonons. Charges diffuse until they either trap (with a probability set by the trapping length $\lambda_{\rm {trap}}$) or until they encounter a surface. Details on tuning of the scattering parameters for this simulation can be found in ref~\citenum{SiChargeTransport}.

We then divide these simulations into bins by initial $z$-position using a bin width of $10~\mu$m. For each of the initial $z$-positions, we compute the probability that the final charge position falls within a bin of width 10~$\mu$m in $x$ and $y$ and 3.71~$\mu$m in $z$ at a given point in a three-dimensional grid the size of the chip under test; with this choice, each dimension has 101 total bins centered at the origin. This bin width allows the PDF to have a resolution much smaller than the lateral extent of the qubits, while the number of bins is large enough to ensure convergence at the tails of the distribution. This set of PDFs over the range of $z$-positions of the impact event is then used to generate final positions of the electrons and holes by random weighted choice.

Using the $\gamma$ and muon tracks derived from the GEANT4 simulations and the charge distributions described above, we generate single- and two-qubit offset charge histograms for a range of values of the characteristic charge trapping length $\lambda_{\rm trap}$ and charge production efficiencies $f_q$. Here, $f_q$ represents the fraction of free charge that avoids immediate recombination at the impact site. In the absence of an applied electric field, we expect $f_q<1$ \cite{Sundqvist:2012wya,Moffatt:2016kok}. To compare our simulations to the measured results, we consider three quantities derived from the charge histograms, described here in order of importance.  The first, \textit{correlation probability} (denoted $p_{ij}^{\rm corr}$ above), is the probability that a discrete jump in offset charge that is registered by one qubit is also registered by its neighbor.  Second, \textit{charge asymmetry} is the number of large positive jumps in offset charge divided by the total number of large jumps, averaged over all four qubits.  While in many cases the change in offset charge measured by the qubit is aliased to the interval from -0.5$e$ to 0.5$e$, no aliasing will occur for a distant source of charge that induces a small change in offset charge $|\Delta q|<0.5e$. We find in our measurements a clear excess of positive offset charge, corresponding to an excess of negative charge in the substrate near the qubit island. We understand the charge asymmetry to arise from the different ways that electrons and holes diffuse in the Si substrate, a consequence of Si valley physics (see discussion above). Finally, we examine \textit{13/24 asymmetry}, the asymmetry in the rate of joint events in quadrants 1 and 3 as opposed to quadrants 2 and 4. We observe a clear excess of events in quadrants 1 and 3, corresponding to distant charge bursts that are not aliased and which couple more or less equally to the two qubits. 

\begin{figure*}[p]
\includegraphics[width=\textwidth]{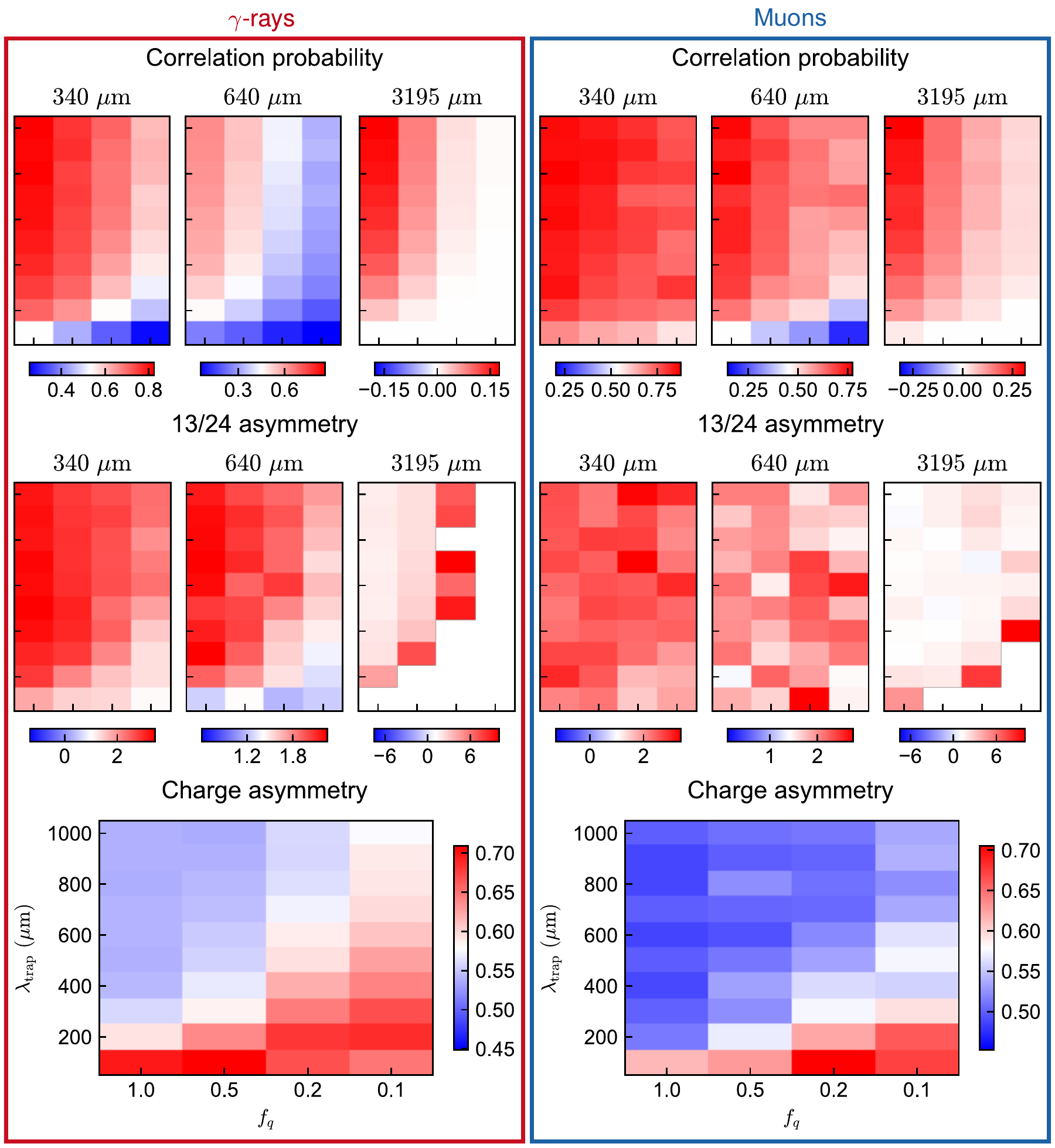}
\caption{ \textbf{Comparison of measured and simulated charge histograms.} Correlation probability, 13/24 asymmetry, and charge asymmetry for simulated joint and single-qubit charge histograms associated with $\gamma$-ray (left) and cosmic ray muon (right) absorption. For correlation probability and 13/24 asymmetry, results are derived from joint histograms calculated for qubits with separation 340~$\mu$m, 640~$\mu$m, and 3195~$\mu$m; for charge asymmetry, results are derived from simulated single-qubit charge histograms. The color scale is adjusted for each plot so that white matches the value derived from the experimentally measured charge histogram. For all plots, the vertical axis corresponds to charge trapping length $\lambda_{\rm trap}$ in the range from 100~$\mu$m to 1000~$\mu$m while the horizontal axis corresponds to charge production efficiency $f_q = 1.0, 0.5, 0.2$, and $0.1$; the axis labels and scales shown in the plots of charge asymmetry apply to all subplots of this figure. For comparison with measurement, we focus on the simulated histograms corresponding to $\gamma$-ray events, which account for over 97\% of the impact events on the qubit chip. From these simulations, we find that the parameter choice $\lambda_\mathrm{trap}=300~\mathrm{\mu m}$ and $f_q=0.2$ provides a good overall match to the experimentally measured data.}
\label{fig:L_fq}
\end{figure*}

\begin{figure*}
\includegraphics[width=\textwidth]{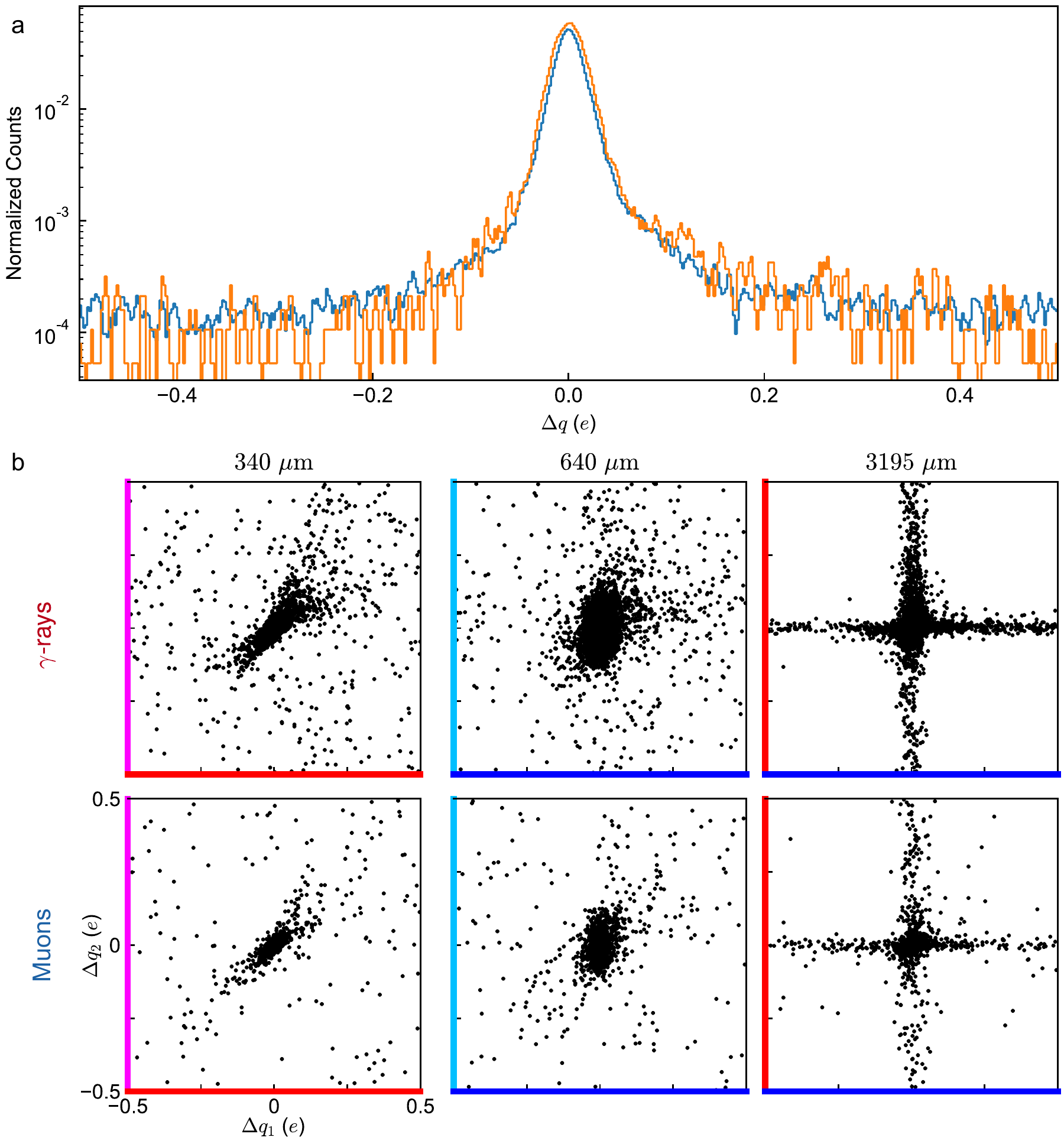}
\caption{ \textbf{Simulated histograms of charge jumps.} (\textbf{a}) Histogram of charge jumps on qubit 1.  The experimentally measured data is shown in blue, while the simulated data is shown in orange.  (\textbf{b}) Simulated joint charge histograms for the three qubit pairs studied in this work. The charge histograms from 7895 $\gamma$-ray events are shown in the top row, while the charge histograms from 1162 cosmic ray muon events are shown in the bottom row.  All simulations were performed with $\lambda_\mathrm{trap}=300$~$\mu$m and $f_q=0.2$ and include Gaussian charge noise comparable to that in the measurements.}
\label{fig:hists}
\end{figure*}

In Fig. \ref{fig:L_fq}, we show extracted values of correlation probability, charge asymmetry, and 13/24 asymmetry for simulations performed for a range of values of $\lambda_\mathrm{trap}$ and $f_q$ for both $\gamma$-ray and muon events.  For all simulated data, we have added Gaussian charge noise derived from the experimental uncertainties associated with Ramsey-based reconstruction of charge. The color scale is set such that white is a match to the value measured experimentally.  As the rate of $\gamma$ impacts on the chip exceeds the muon rate by roughly a factor 40, we focus on simulated $\gamma$-ray events for the purposes of comparison with the measured data. While simulations performed at large $\lambda_{\rm trap}$ and small $f_q$ yield qualitatively similar results for correlation probability, charge asymmetry, and 13/24 asymmetry as simulations performed at small $\lambda_{\rm trap}$ and large $f_q$, the degeneracy with respect to the choice of parameters can be broken by examining the detailed shape of the single- and two-qubit charge histograms. We find that the parameter choice $\lambda_\mathrm{trap}=300$~$\mu$m and $f_q=0.2$ provides the best overall agreement with the measured data. This implies a low mean charge collection, consistent with zero-field measurements in comparable crystals \cite{Moffatt:2016kok}.

In Fig. \ref{fig:hists}a, we display the single-qubit charge histogram derived from the simulated $\gamma$-ray events along with the measured histogram of discrete charge jumps on Q1. In Fig. \ref{fig:hists}b, we show the simulated two-qubit histograms for the $\gamma$-ray and muon events. Despite good agreement with the measured joint charge histograms across correlation probability, charge asymmetry, and 13/24 asymmetry, the simulated joint histogram for the smallest qubit separation of 340~$\mu$m displays a pronounced charge ``jet" in quadrants 1 and 3. This feature arises from distant charge bursts that couple equally to the two qubits. We believe that in the experimental system this feature will be suppressed, as field lines from faraway bursts will close on the surrounding metal of the qubit enclosure, providing a natural cutoff in the response to distant charge. More sophisticated modeling of the qubit setup, including the detailed geometry of the sample enclosure, could capture this physics; however, this is beyond the scope of the current work.

\subsection{Implications for Fault Tolerance}

\begin{figure}
\includegraphics[width=\columnwidth]{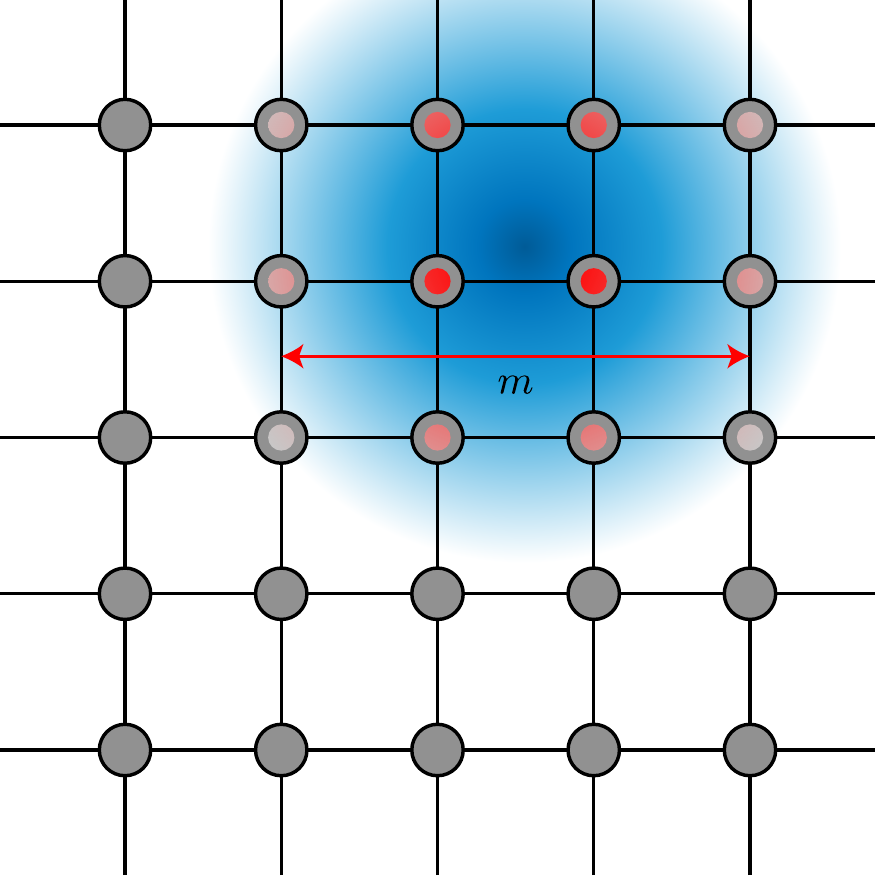}
\caption{ \textbf{Correlated errors in the surface code.} Schematic of a surface code array subjected to a correlated noise source, shown here in blue. The number $m$ of qubits in a line to which the noise couples determines the sensitivity of the array to correlated errors. }
\label{fig:fault}
\end{figure}

Here we briefly discuss the impact of correlated errors on error identification in the surface code; for a complete analysis, see ref. \citenum{Faoro20}. The surface code consists of a two-dimensional fabric of qubits with nearest-neighbor connectivity. We define correlation degree $m$ as the number of qubits in a line to which an error couples (see Fig.~\ref{fig:fault}). It can be shown that the fault-tolerant threshold $p_m$ for correlated errors of degree $m$ is given by
\begin{equation}
    p_m \approx p^m,
\end{equation}
where $p$ is the fault-tolerant threshold for uncorrelated errors. Thus, for a threshold error level $p = 10^{-2}$, the threshold for correlated errors of degree 2 is $p_2 \approx 10^{-4}$; the threshold for correlated errors of degree 3 is $p_3 \approx 10^{-6}$, etc. These considerations must inform the design of large-scale qubit arrays that are susceptible to correlated errors: for a given error mechanism with a fixed spatial footprint, the need to protect against correlated errors will dictate how closely spaced the qubits can be.

Due to the exponential dependence of the fault-tolerant threshold on correlation degree, errors due to widespread quasiparticle poisoning are particularly damaging. In the following, we separately analyze in detail two additional correlated error mechanisms, in order of importance: correlated phase errors due to exponentially small (but nonzero) frequency shifts induced by correlated charge noise, and correlated bit-flip errors induced by pair production and charge reconfiguration in the substrate at the moment of particle impact.

\subsection{Qubit Phase Flips From Correlated Charge Noise}

\begin{figure*}[p]
\includegraphics[width=\textwidth]{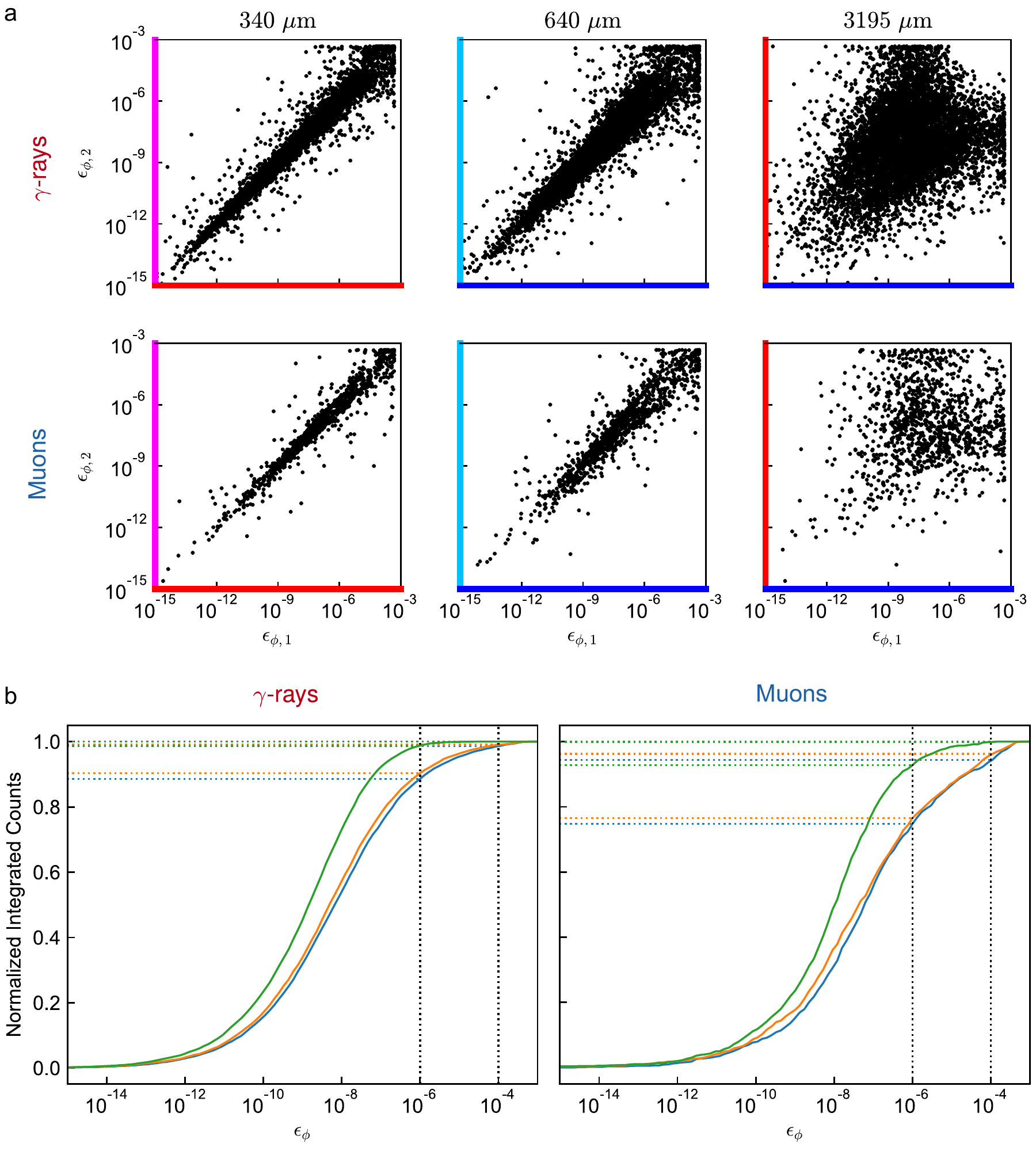}
\caption{ \textbf{Phase-flip errors from correlated charge fluctuations.} (\textbf{a}) Histograms of joint phase-flip errors from correlated charge noise induced by particle impacts. From left to right, the plots correspond to qubit pairs with center-to-center separation 340~$\mu$m, 640~$\mu$m, and 3195~$\mu$m; the top row corresponds to 7895 simulated $\gamma$-ray events, while the bottom row corresponds to 1162 simulated muon events. The qubit geometry and chip layout are identical to those considered throughout this work, but we take the more typical transmon parameters $E_J/h$~=~12.5~GHz and $E_C$~=~250~MHz. (\textbf{b}) Integrated histogram of correlated phase-flip errors from charge fluctuations induced by $\gamma$-ray (left) and muon (right) impacts. Blue, orange, and green traces correspond to qubit pairs with center-to-center separation 340~$\mu$m, 640~$\mu$m, and 3195~$\mu$m, respectively.}
\label{fig:phase_err}
\end{figure*}

The devices used in this work were intentionally designed to be sensitive to charge. For a conventional transmon qubit optimized for high-fidelity gates, the sensitivity to charge is exponentially small, and charge noise has negligible impact on device coherence. However, due to the exponential sensitivity of the qubit array to correlated errors, it is necessary to carefully examine errors due to charge fluctuations that are sensed by multiple qubits. For a transmon qubit, the charge dispersion of the 01 transition $\Delta \omega_{01}$ (defined as half the peak-to-peak value) can be written as \cite{JKoch07}
\begin{equation}
       \Delta \omega_{01} = 16 \sqrt{\frac{2}{\pi}} \frac{E_C}{\hbar} \left(\frac{\xi}{2}\right)^{3/4} e^{-\sqrt{8\xi}}\left[16\left(\frac{\xi}{2}\right)^{1/2} + 1 \right],
\end{equation}
where we have defined $\xi \equiv E_J/E_C$. The qubit transition frequency depends on charge as follows:
\begin{equation}
    \omega_{01} = \overline{\omega_{01}} - \Delta \omega_{01} \cos \left[\frac{\pi}{e}\left(q_0 + \Delta q\right)\right],
\end{equation}
where $\overline{\omega_{01}}$ is the mean qubit transition frequency, $q_0$ is a random offset charge, and $\Delta q$ is a discrete change in offset charge due to particle impact in the qubit substrate. The mean-square frequency shift associated with the charge jump $\Delta q$ is given by 
\begin{equation}
    \expect{\delta \omega_{01}^2} = 2 \Delta \omega_{01}^2 \sin^2 \left( \pi \frac{\Delta q}{2e} \right),
\end{equation}
where we have averaged over the random offset charge $q_0$. We can convert the frequency shift to a phase error accumulated during a surface-code cycle time $\tau_{\rm sc}$ as follows: 
\begin{equation}
    \expect{\phi^2} = \expect{\delta \omega_{01}^2} \tau_{\rm sc}^2.
\end{equation}
Averaging over qubit states aligned along the six cardinal directions of the Bloch sphere, we find a phase-flip error probability $\epsilon_\phi$ associated with charge jumps that is given by 
\begin{equation}
    \epsilon_\phi = \frac{\Delta \omega_{01}^2 \tau_{\rm sc}^2}{3} \sin^2 \left( \pi \frac{\Delta q}{2e} \right).
    \label{eqn:phase_err}
\end{equation}
In Fig.~\ref{fig:phase_err}a we plot histograms of two-qubit phase-flip errors derived from the simulations presented in Fig.~\ref{fig:hists}. Here we take $\tau_{\rm sc}$~=~1~$\mu$s and we take conventional transmon parameters $E_C/h$~=~250~MHz and $E_J/h$~=12.5~GHz, corresponding to a mode frequency around 5~GHz and a charge dispersion $\Delta \omega_{01}/2\pi$~=~6.0~kHz. It is important to note that while our Ramsey-based charge measurement aliases charge fluctuations into the interval from -0.5$e$ to +0.5$e$, for the purposes of dephasing the qubit will be sensitive to charge fluctuations in the interval from $-e$ to $+e$, as the surface code cycle time $\tau_{\rm sc}$ is short compared the characteristic quasiparticle parity dwell time.  In Fig.~\ref{fig:phase_err}b we plot integrated histograms of correlated phase-flip errors. We find that 11\% (1.3\%) and 9.7\% (0.99\%) of $\gamma$-ray events induce correlated errors above the $10^{-6}$ ($10^{-4}$) level in qubit pairs with separation 340~$\mu$m and 640~$\mu$m, respectively. For muon absorption, 25\% (5.6\%) and 23\% (3.8\%) of events induce simultaneous errors above the $10^{-6}$ ($10^{-4}$) level in qubit pairs with separation 340~$\mu$m and 640~$\mu$m, respectively. Finally, 7.2\% of muon events induce correlated phase-flip errors above the $10^{-6}$ level for qubits separated by 3195~$\mu$m. 

Based on the analysis presented above, correlated charge fluctuations can be a significant error mechanism for surface-code arrays implemented using the transmon qubit. The sensitivity of the transmon to charge noise can be further suppressed by increasing the ratio $E_J/E_C$; for fixed mode frequency, however, such an approach leads to reduced anharmonicity $\omega_{01} - \omega_{12} \approx E_C/\hbar$ and increased leakage errors for fixed gate times. The need to protect against correlated phase-flip errors due to charge noise from particle impacts thus represents an important constraint to the design of fault-tolerant transmon qubit arrays.

\subsection{Qubit Bit Flips From Charge Bursts}

\begin{figure*}
\includegraphics[width=\textwidth]{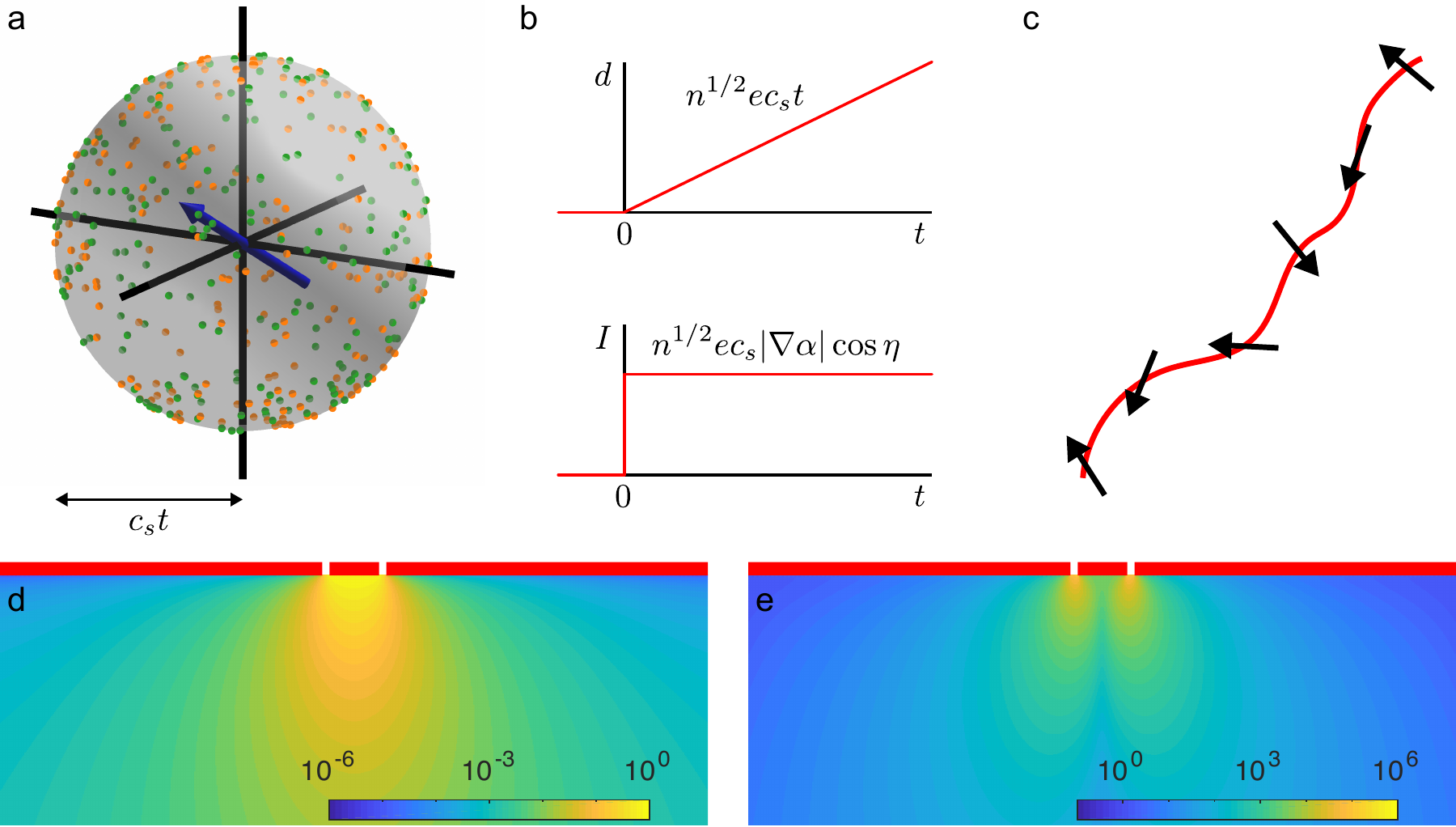}
\caption{ \textbf{Qubit response to nonadiabatic charge bursts.} (\textbf{a}) As electrons (green) and holes (orange) move ballistically with speed $c_s$ from the site of pair production, the randomly oriented electric dipole moment associated with the charge distribution grows linearly with time $t$ following the impact event. (\textbf{b}) Dipole moment $d$ and effective current drive $I$ to a neighboring qubit \textit{versus} $t$. (\textbf{c}) The net response of the qubit to the nonadiabatic charge shifts associated with particle absorption in the substrate can be obtained by summing over the responses to the random electric dipoles that are nucleated along the particle track. (\textbf{d}) Dimensionless offset charge $\alpha(\bm{r})$ induced on the qubit island by a unit charge at point $\bm r$. Vertical extent of the plot corresponds to the 375~$\mu$m thickness of the chip, while the horizontal extent corresponds to 2~mm centered on the qubit island (not to scale). (\textbf{e}) Gradient $|\bm{\nabla} \alpha(\bm{r})|$, in units of m$^{-1}$.}
\label{fig:RD} 
\end{figure*}

\begin{figure*}
 \includegraphics[width=\textwidth]{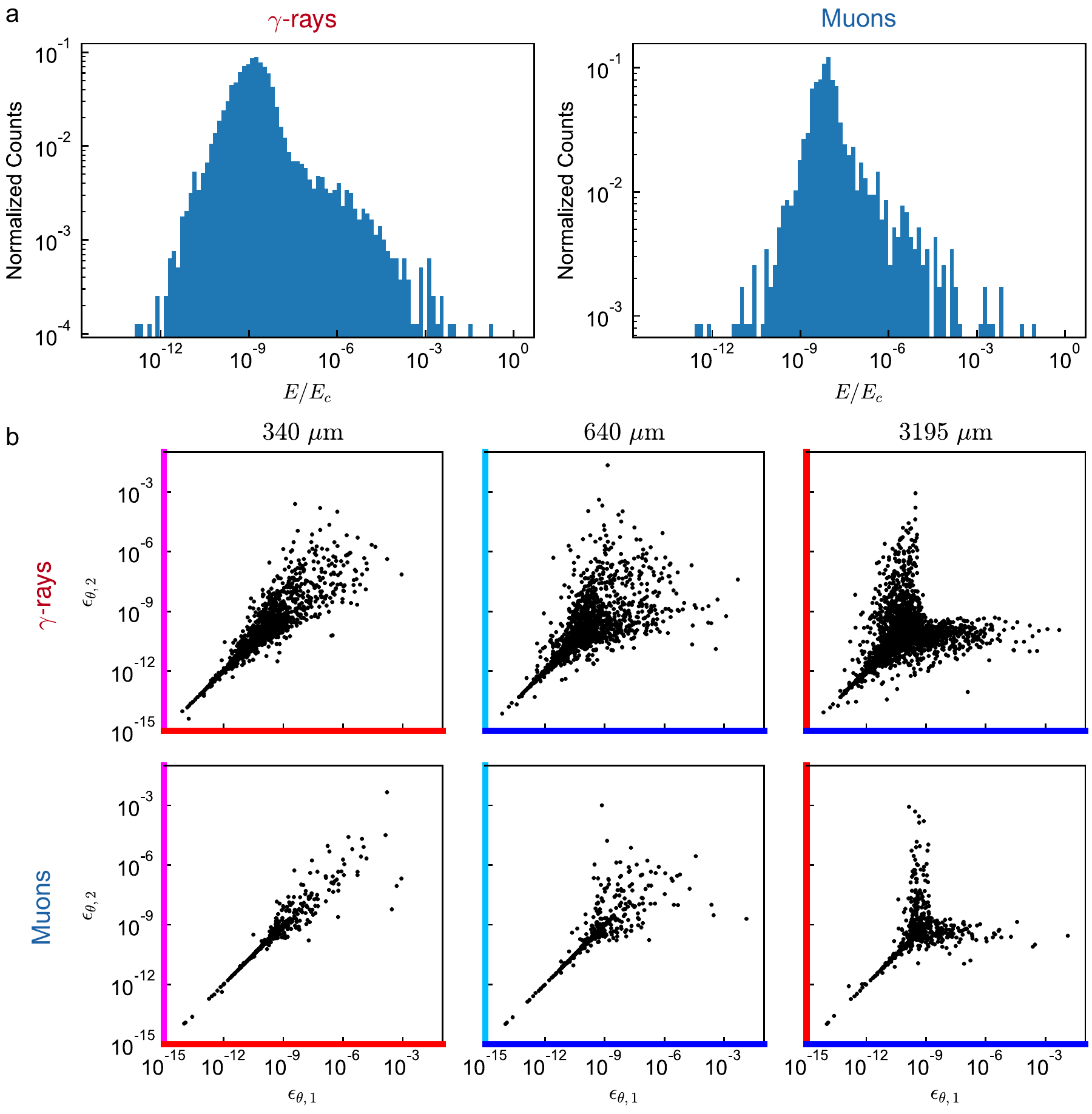}
 \caption{ \textbf{Bit-flip errors from nonadiabatic charge bursts.} (\textbf{a}) Histogram of energy $E$ deposited in the qubit mode by the charge transient associated with particle impact in the substrate. Energy is expressed in terms of the single-electron charging energy $E_C$. The plot on the left is calculated from 7895 simulated $\gamma$-ray absorption events, while the plot on the right is calculated from 1162 simulated muon absorption events. The qubit geometry and chip layout are identical to those considered throughout this work. (\textbf{b}) Histograms of joint error associated with the charge transients. From left to right, the plots correspond to qubit pairs with center-to-center separation 340~$\mu$m, 640~$\mu$m, and 3195~$\mu$m; the top row corresponds to $\gamma$-ray events, while the bottom row corresponds to muon events.}
 \label{fig:err}
 \end{figure*}
 
We identify two mechanisms for spurious qubit bit flips due to nonadiabatic reconfiguration of charge in the substrate following particle absorption. The first is associated with the sudden release of charge in the immediate aftermath of the impact, while the second is associated with single discrete charges that propagate from the impact site all the way to the qubit island. We discuss the relevant physics below. 

The charges liberated by energy absorption in the substrate at point $\bm{r}$ very quickly emit phonons and relax to an energy where further phonon emission is kinematically forbidden; at this point, electrons and holes are moving ballistically at $c_s$, the sound speed in the substrate. We can fully characterize any charge distribution by specifying its multipole moments; the highest nonvanishing multipole moment will dominate the coupling to distant qubits. The electrons and holes liberated by the impact preserve charge neutrality; therefore, the monopole moment (net charge) is zero. There will, however, be a nonvanishing random dipole moment $\bm{d}(\bm{r})$ that grows linearly in time $t$ with respect to the moment of impact:
\begin{equation}
d(t) = n^{1/2} e \, c_s t,	
\end{equation}
where $n$ is the number of discrete charges liberated by the impact event (see Fig.~\ref{fig:RD}). This random dipole will induce an offset charge on the qubit island given by 
\begin{equation}
Q(t) = \bm{d} (\bm{r},t) \cdot \bm{\nabla} \alpha (\bm{r}),	
\end{equation}
where $\alpha(\bm{r})$ is the offset charge induced on the qubit island by a unit point charge at location $\bm{r}$. Note that the time dependence of $Q(t)$ is the same as that of $d(t)$: we have $Q(t) = 0$ for $t<0$, and $Q(t) \propto t$ for $t>0$. The abrupt transient at the moment of impact $t=0$ is what drives qubit transitions. This transient corresponds to a current step:
\begin{equation}
I(t) = n^{1/2} e c_s \left|\bm{\nabla} \alpha \right| \cos \eta \,  H (t),	
\end{equation}
where $\eta$ is the angle between $\bm{\nabla} \alpha$ and the random dipole $\bm{d}$ and $H(t)$ is the Heaviside step function. This transient current deposits an energy in the qubit given by 
\begin{eqnarray}
E(\bm{r}) &=& \frac{1}{2C}\left|\tilde{I}(\omega_{01})\right|^2 \nonumber \\
&=& \frac{n c_s^2 \left|\bm{\nabla} \alpha \right|^2}{\omega_{01}^2}\, E_C\, \cos^2 \eta,	
\end{eqnarray}
where $C$ is the self-capacitance of the qubit and $\tilde{I}(\omega_{01}) = \int_{-\infty}^{+\infty} I(t) e^{-i\omega_{01}t} \, dt$ is the Fourier transform of the current step evaluated at the qubit frequency. 

We can equivalently express the coupling of the random dipole to the qubit in terms of a spurious rotation angle $\theta$ using the relation 
\begin{equation}
E(\bm{r}) \approx \hbar \omega_{01} \, \frac{\theta^2}{4}.	
\end{equation}
We find
\begin{equation}
\theta(\bm{r}) = 2 n^{1/2}c_s \left|\bm{\nabla} \alpha \right|\left(\frac{E_C}{\hbar\omega_{01}^3}\right)^{1/2}\, \cos \eta.	
\end{equation}

Similarly, we can express the coupling of the random dipole to the qubit as a bit-flip error probability $\epsilon_\theta = \theta^2/6$, where we have performed an average of the qubit response to the charge burst over qubit states aligned along the six cardinal directions on the Bloch sphere. We find
\begin{equation}
\epsilon_\theta(\bm{r}) = \frac{2}{3} n c_s^2 \left|\bm{\nabla} \alpha \right|^2 \frac{E_C}{\hbar\omega_{01}^3} \cos^2 \eta.	
\end{equation}

For a $\gamma$-ray or muon track, we expect to have energy deposited over a range of points in the substrate at times that are short compared to the qubit oscillation period. In this case, we need to add the rotations induced by the separate energy deposits. Using this model, we have calculated single-qubit and joint error probabilities associated with $\gamma$-ray and muon absorption in the substrate. In Fig.~\ref{fig:err}a we plot the total energy deposited in the qubit in terms of $E_C$ for $\gamma$-ray and muon tracks; here, we use the $\gamma$-ray and muon tracks calculated using GEANT4 as described above, the electrostatic response function $\alpha(\bm{r})$ calculated for our geometry, and we take the typical qubit frequency $\omega_{01}/2 \pi$ = 5~GHz. In Fig.\ref{fig:err}b, we plot joint errors for this set of $\gamma$-ray and muon tracks. Again we take a qubit frequency of 5~GHz and we take a charging energy $E_C/h$~=~250~MHz. We find that 1.2\% and 0.7\% of $\gamma$-ray events induce simultaneous errors above the $10^{-8}$ level in qubit pairs with separation 340~$\mu$m and 640~$\mu$m, respectively. For muon absorption, 4.0\% and 3.1\% of events induce simultaneous errors above the $10^{-8}$ level in qubit pairs with separation 340~$\mu$m and 640~$\mu$m, respectively.

Following the initial transient associated with particle absorption, a charge-insensitive qubit will follow the slow drift of distant charge in the substrate adiabatically. However, it is possible that a charge that propagates all the way to the qubit island will give rise to a nonadiabatic shift in island charge as it passes through a region where $\left|\bm{\nabla}\alpha(\bm{r})\right|$ is large. The crossover from adiabaticity to nonadiabaticity can be expressed as 
\begin{eqnarray}
\frac{d\omega_{01}}{dt} &\sim& \omega_{01}^2 \nonumber \\
\frac{c_s \left|\bm{\nabla} \alpha(\bm{r})\right|}{\hbar} \, E_C &\sim& \omega_{01}^2.
\end{eqnarray}
This relation defines a surface surrounding the qubit island; charges that cross this surface will induce a sudden, nonadiabatic shift in offset charge, resulting in an error of order $E_C/\hbar \omega_{01}$. For the electrostatic response function $\alpha(\bm{r})$ associated with our qubit geometry, we find that rotation errors induced by direct charge impingement on the qubit island will be negligible; however, this error mechanism could be important for other device geometries or parameters.

\subsection{Characterization of Background Radiation}
\label{sec:bkg_measurement}

\begin{figure}
\includegraphics[width=\columnwidth]{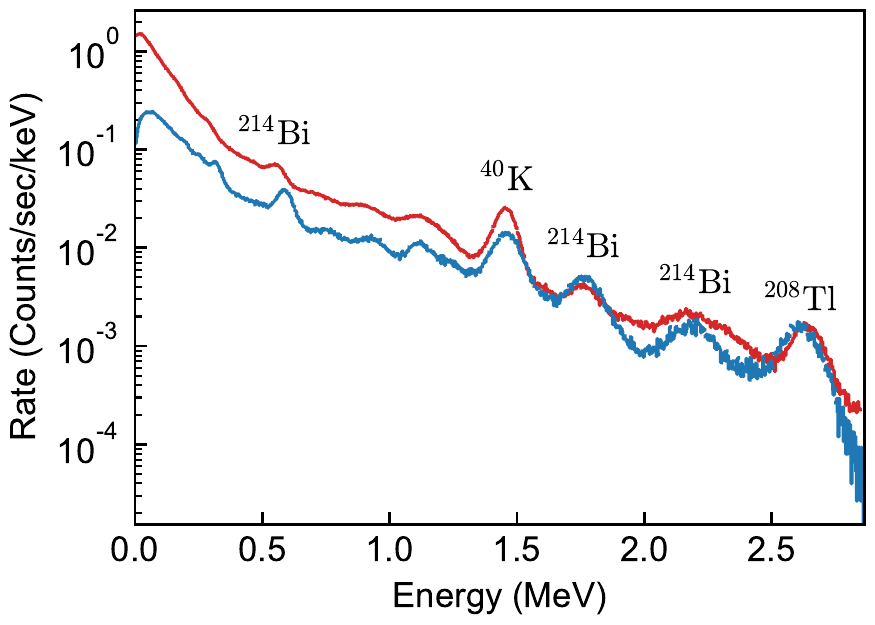}
\caption{ \textbf{Spectrum of background radioactivity.} Red: spectrum of environmental $\gamma$ radiation measured in the laboratory in Madison with a 1.5'' NaI scintillation detector.  Blue: spectrum of environmental $\gamma$ radiation measured in LNGS with a 3'' NaI scintillation detector and used for the simulations (see Sec.~\ref{sec:modeling}). The Madison spectrum has been scaled by a factor 5.2 to account for the smaller detector dimension \cite{Courbois1969}, allowing a direct comparison with the spectrum from LNGS. 
The distinguishable peaks of $^{40}$K at 1.460\,MeV and $^{208}$Tl at 2.614\,MeV can be seen as dominant contributors to the spectra. The integrated $\gamma$-ray flux measured in Madison is a factor 2.8 larger than that measured at LNGS.}
\label{fig:NaI Energy Spectrum}
\end{figure}

We use a 1.5'' NaI scintillation detector to characterize the spectrum of background radioactivity in the qubit laboratory in Madison. We use sealed sources of $^{137}$Cs and $^{60}$Co with known activity to calibrate the 
detector at the photopeak energies 662~keV, 1.17~MeV, and 1.33~MeV. To bootstrap the calibration to higher energy, we use the 1.37 and 2.75~MeV $\gamma$-ray emission from $^{24}$Na obtained by proton irradiation of a piece of Al in the UW-Madison Cyclotron Laboratory. In Fig.~\ref{fig:NaI Energy Spectrum} we plot the measured $\gamma$-ray spectrum, along with the $\gamma$-ray spectrum measured with a 3'' NaI detector in the underground laboratory at LNGS in Gran Sasso, which was used as input to the $\gamma$-ray simulations performed using GEANT4. Here the spectrum measured in Madison has been scaled by a factor 5.2 to account for the smaller detector dimension \cite{Courbois1969}, facilitating direct comparison with the spectrum from LNGS. Apart from a factor 2.8 difference in absolute rate, the spectra are similar. The background activity is dominated by the 1.46~MeV $\gamma$ emission of $^{40}$K and the 2.61~MeV $\gamma$ emission of $^{208}$Tl, both of which are known to be common radioactive contaminants. Again, the spectrum of energy deposited in the qubit chip is insensitive to the fine details of the spectrum of background radiation used for the simulations. 

From the $\gamma$-ray spectrum measured in Madison and the GEANT4 simulations, we expect a rate of $\gamma$ impacts on the qubit chip of 9~mHz. As we discuss above, from our measured rate of offset charge jumps, we infer a rate of $\gamma$ impacts on the chip of 19.8(5)~mHz. The factor 2 discrepancy could indicate a dominant local source of radioactivity within the qubit cryostat: the qubit chip is closely surrounded by copper stage plates, mu-metal shields, stainless steel vacuum cans, etc., all of which are potential emitters of low-level $\gamma$ radiation. A systematic study of the various contributions to the $\gamma$-ray impact rate in the qubit chip will the the focus of future work.


\end{document}